\pdfoutput=1
\documentclass{ws-ijmpc}
\usepackage{epsfig}

\usepackage{amsmath}

\newcommand{\D}{\ensuremath{{\mathrm{d}}}}
\newcommand{\imag}{\ensuremath{{\mathrm{i}}}}
\newcommand{\E}{\ensuremath{{\mathrm{e}}}}

\begin{document}

\title{COMPARATIVE STUDY OF SEMICLASSICAL APPROACHES TO QUANTUM DYNAMICS}

\author{ G.~SCHUBERT}
\address{Institut f\"ur Physik, Ernst-Moritz-Arndt Universit\"at Greifswald\\
  17487 Greifswald, Germany\\
  schubert@physik.uni-greifswald.de}

\author{V.S.~FILINOV}
\address{ Joint Institute for High Temperatures, Russian Academy of Sciences\\
  Moscow 127412, Russia}

\author{K.~MATYASH, R.~SCHNEIDER}
\address{Max-Planck-Institut f\"ur Plasmaphysik\\
  17491 Greifswald, Germany}

\author{H.~FEHSKE}
\address{Institut f\"ur Physik, Ernst-Moritz-Arndt Universit\"at Greifswald\\
  17487 Greifswald, Germany}

\DeclareGraphicsExtensions{.pdf}
\graphicspath{{./eps/}}

\maketitle


\begin{abstract}
  Quantum states can be described equivalently by density matrices, 
  Wigner functions or quantum tomograms.
  We analyze the accuracy and performance of three related semiclassical 
  approaches to quantum dynamics, in particular with respect to their 
  numerical implementation. 
  As test cases, we consider the time evolution of Gaussian wave
  packets in different one-dimensional geometries, whereby tunneling, 
  resonance and anharmonicity effects are taken into account.
  The results and methods are benchmarked against 
  an exact quantum mechanical treatment of the system, which is 
  based on a highly efficient Chebyshev expansion technique 
  of the time evolution operator.
\end{abstract}

\ccode{PACS Nos.: 03.65.Sq, 
  03.65.Wj}  

\footnotetext{\hrule\vspace*{2mm}
Electronic version of an article published as DOI:10.1142/S0129183109014278\\
 {\it International Journal of Modern Physics C},
 Vol. 20, No. 8 (2009) pp. 1155 - 1186,\\
$\copyright$ World Scientific Publishing Company, http://www.worldscinet.com/ijmpc/ijmpc.shtml}

\section*{Introduction}

Quantum statistical physics, such as condensed matter or plasma physics,
but also quantum chemistry, heavily depends on effective 
numerical methods for solving complex few- and many particle problems.
Implementing suitable theoretical concepts for their description
on modern (super-) computer architectures, nowadays computational physics 
constitutes, besides experiment and theory, the third column of 
contemporary physics~\cite{FWS08}. 
Numerical techniques become especially important for strongly correlated systems 
where analytical approaches largely fail.
This may be due to the absence of small (coupling) parameters or, more general, 
because the relevant energy scales are not well separated, both preventing 
the application of standard perturbative schemes. 
Another challenging problem,
that calls for numerical approaches, concerns the description of 
microscopic and nanoscopic systems, and of particles in finite quantum 
structures (restricted geometries), where the level quantization might 
become as important as particle correlations.

A first idea for the numerical study of this kind of quantum systems might be a  
brute force exact diagonalization of the underlying (model) Hamiltonian.
Considering the exponential growth of the Hilbert space with the number
of particles, such a description of quantum many-particle systems is 
(and will be in the future) out of reach.
The situation becomes even more difficult if bosonic degrees of freedom 
(e.g. phonons in a solid) come into play, resulting in an infinite dimensional
Hilbert space even for finite systems. 
One way to circumvent this problem is to restrict the many-particle Hilbert 
space to the physically most important subset. 
Along this line, e.g., density matrix 
renormalization group schemes have been developed~\cite{Pe99,JF07}.
Semiclassical descriptions offer another possibility to overcome
these limitations.
A variety of semiclassical methods has been
proposed during the last decades, especially to describe the dynamics of
quantum systems~\cite{Mi01,TW04,Mi70,BM72,He81,Ka05}. 
These methods are appealing, in some sense, as their close 
relation to concepts familiar from classical physics 
facilitates an intuitive interpretation of the results.
Also in view of bridging the gap between quantum many-particle 
and classical continuum theories they seem promising for describing 
systems in the thermodynamic limit. 
Therefore we will focus on semiclassical approaches to quantum dynamics
in the following, of which we will discuss three carefully selected ones in 
more detail.

The majority of the traditional semiclassical methods is based
on a real time path integral formulation of 
quantum mechanics~\cite{Fe48,FH65}. 
Expressing the time evolution of the complex wave function in terms of 
action integrals, the Feynman path integral is generally evaluated 
within the stationary-phase approximation. 
Then the occurring integrals can sometimes be performed analytically,
otherwise numerically using direct integration or Monte Carlo (MC) 
techniques~\cite{Fi86}. 
Integrating an oscillatory, complex valued integrand, the
dynamical sign problem, however, spoils to some extend the efficiency
of the MC integration.
Despite the exponential decay of the integrand outside a vicinity of 
the classical trajectories the numerics is still challenging.
In the sum of many contributions with different complex phases most 
of them may cancel out and the result may become exponentially small.

Equivalently, a quantum system can be described in terms of 
real valued quantum phase space distribution functions~\cite{Le95},
e.g., the Wigner function~\cite{Wi32}.
This overcomes the problem of handling a complex valued wave function
but, because of possible negative values of the Wigner function 
and the Heisenberg uncertainty relation, the Wigner 
function cannot be interpreted as a joint probability.
Instead, it should be considered as a convenient mathematical tool 
for the description of quantum systems. In the numerical work, 
the dynamical sign problem is alleviated
for the Wigner function but still present~\cite{FMK95,Fi96c,FBFG08}.

In order to overcome the heavily debated dynamical sign problem, 
some years ago the description of quantum states in terms of a 
positive function, the so called quantum tomogram, has been 
proposed~\cite{MMT96,MMT97}.
The strict positivity of the tomogram seems promising in view of an 
effective MC sampling of the trajectories. 
In the framework of the tomographic representation a description of 
quantum dynamics by diffusive Markov processes was 
suggested in~\cite{FSLBFFF08}.
Recently, the applicability of the tomographic approach to 
one- and two-particle systems has been demonstrated~\cite{ALM04,LSA04}. 
However, to the best of our knowledge, most of those studies are 
based on harmonic potentials but validations for arbitrarily 
shaped potentials are still missing.

Motivated by this situation, it is the aim of the present work, 
to outline and compare the various methods quoted
above with respect to their accuracy and computational performance.
The selected semiclassical methods were chosen to represent algorithms
based on different descriptions of quantum states:
the wave function, the Wigner function and the quantum tomogram.
Instead of investigating complex physical systems in which a multitude
of effects compete, we benchmark the methods on the basis of relatively 
simple systems, concentrating onto a single aspect in each case. 
We will focus on the dynamics of a wave packet in distinct model
geometries which account for basic aspects of quantum mechanics: 
tunneling, confinement and nonlinearity (anharmonicity).
Calibrating the different approaches by studying simple toy models is necessary 
in order to detect their limitations and prospects, before applying them
to the more complicated problems of current interest, e.g., to 
classical chaotic systems, quantum localisation, dynamic tunneling or entanglement.
The obtained semiclassical results will be compared with
an exact solution for the time evolution of the quantum system.
Thereby, the exact solution is calculated using a highly efficient
expansion of the time evolution operation in a series of Chebyshev
polynomials.
%


\section{Computational schemes}
\label{methods}

\subsection{Chebyshev expansion of the time evolution operator}

As a reference for the approximate results we will present in the next
sections, the Chebyshev
expansion of the time evolution operator constitutes a very efficient 
technique which fully includes all quantum effects.
Governed by the time dependent Schr{\"o}dinger equation, the dynamics of 
a quantum state $|\psi(t_0)\rangle$ in a time independent external potential may
be expressed in terms of the time evolution operator $U(t,t_0) = U(\Delta t) =
\E^{ -\imag H (t-t_0)/\hbar}$ with $\Delta t=t-t_0$.
Expanding $U(\Delta t)$ into a series of first kind Chebyshev polynomials
of order $k$,
$T_k(x)=\cos(k\, \mathrm{arccos}(x))$,~\cite{TK84,CG99,WF08},
we obtain
\begin{equation}
  U(\Delta t) =  \E^{-\imag b \Delta t/\hbar} 
  \left[ c_0(a\Delta t/\hbar) + 2\sum\limits_{k=1}^{M} c_k(a\Delta t/\hbar)
    T_k(\tilde{H}) \right]\;.
\label{U_1}
\end{equation}
Prior to the expansion, the Hamiltonian has to be shifted and rescaled such that
the spectrum of $\tilde{H} = (H-b)/a$ is within the definition interval of the
Chebyshev polynomials, $[-1,1]$.
The parameters $a$ and $b$ are calculated from the extreme eigenvalues of $H$ as
$b=\frac{1}{2}(E_{\mathrm{max}}+E_{\mathrm{min}})$ and 
$a=\frac{1}{2}(E_{\mathrm{max}}-E_{\mathrm{min}}+\epsilon)$.
Here, we introduced $\epsilon=\alpha(E_{\mathrm{max}}-E_{\mathrm{min}})$ to ensure 
the rescaled spectrum $|\tilde{E}| \le 1/(1+\alpha)$ to be well inside $[-1,1]$.
In practice, we use $\alpha=0.01$. 
Note, that the Chebyshev expansion also applies to systems with unbounded spectra.
In those cases we truncate the infinite Hilbert space to a finite 
dimension by restricting the model on a discrete space grid or using an energy
cutoff.
By this, we ensure possibly large but finite extreme eigenvalues.

In~(\ref{U_1}), the expansion coefficients $c_k(a\Delta t/\hbar)$ are given by 
\begin{equation}
  c_k(a\Delta t/\hbar) = \int\limits_{-1}^1 
  \frac{T_k(x)e^{-\imag x a \Delta t/\hbar }}{\pi \sqrt{1-x^2}}\D x =
  (-\imag)^k J_k(a \Delta t/\hbar)\;,
\end{equation}
where $J_k$ denotes the $k$-th order Bessel function of the first kind. 

To calculate the evolution of a state $|\psi(t_0)\rangle$ from one 
time grid point to the next one,
$|\psi(t)\rangle = U(\Delta t)|\psi(t_0)\rangle$, we
have to accumulate the $c_k$-weighted vectors
$|v_k\rangle = T_k(\tilde{H})|\psi(t_0)\rangle$. 
Since the coefficients $c_k(a\Delta t/\hbar)$ depend on the time step but not
on time explicitly, we need to calculate them only once.
The vectors $|v_k\rangle$ can then be calculated iteratively using
the recurrence relation of the Chebyshev polynomials
\begin{equation}
    |v_{k+1}\rangle = 2\tilde{H} |v_k\rangle - |v_{k-1}\rangle\;,
\end{equation}
where $|v_1\rangle = \tilde{H} |v_0\rangle$ and $|v_0\rangle = |\psi(t_0)\rangle$.

In the numerics, we use a discrete coordinate space basis $|q_i\rangle$,
$i=1,\ldots,N$ with $\langle q|q_i\rangle = \delta(q-q_i)$, representing an
equally spaced grid of $N$ points.
The wave function at position $q_i$ is given by the $i$-th entry of the corresponding
(complex) vector, $\psi(q_i,t_0) = \langle q_i|\psi(t_0)\rangle$.
Aiming at the description of a spatially unbounded system, we choose
the extension of our space grid such that $\psi(q_1,t) = \psi(q_N,t)\equiv 0$
throughout the whole simulation.
In this way, no artificial reflections at the boundaries of the simulation volume
arise and the results are independent of the actual size of the simulation 
volume.
Overall, evolving the wave function from one time step to the next
requires $M$ matrix vector multiplications (MVM) of a given complex vector with 
the (sparse) Hamiltonian of dimension $N$ as well as the summation of the
resulting vectors after appropriate scaling:
\begin{equation}
  |\psi(t)\rangle =  \E^{-\imag b \Delta t/\hbar} 
  \left[ J_0(a\Delta t/\hbar) + 2\sum\limits_{k=1}^{M} (-\imag)^k J_k(a\Delta t/\hbar) 
    T_k(\tilde{H}) \right]|\psi(t_0)\rangle\,.
\end{equation}
Note that the Chebyshev expansion may also be applied to systems with
time dependent Hamiltonians. 
However, there the time variation $H(t)$ determines the maximum $\Delta t$ by which
the system may be propagated in one time step.
For time independent $H$, in principle, arbitrary large time steps are
possible at the expense of increasing $M$.


\subsection{Linearized semiclassical propagator method}
\label{sect:QPIC}
\begin{figure}
  \centering
  \includegraphics[width=\linewidth,clip]{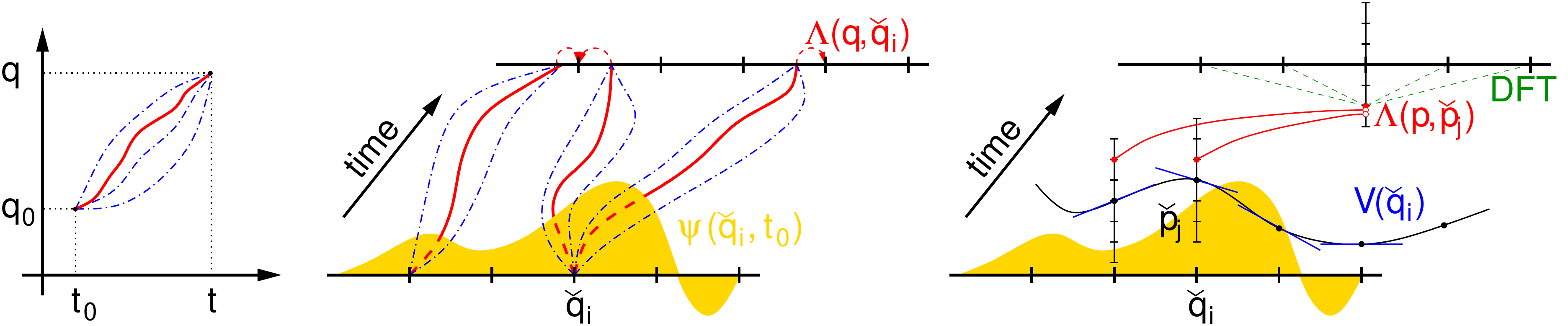}
  \caption{Left panel: Virtual (blue, dashed-dotted lines) and classical
    (red, solid line) paths connecting $q_0(t_0)$ and $q(t)$. For the
    classical path the action along the trajectory is extremal.
    Middle panel: Visualization of the deposition scheme (\ref{eq:prop_grid}).
    At each coordinate grid point $\check{q}_i$ the initial wave function
    $\psi(\check{q}_i,t_0)$ determines the sampling weight for all trajectories 
    starting from $\check{q}_i$. The path propagation follows the classical
    equations of motion and the virtual (non-extremal) paths are taken
    into account perturbatively. As the end points of the paths $q(t)$ need
    not match the grid, they are deposited there using the shape function
    $\Lambda(q,\check{q}_i)$. Right panel: Wave function propagation as 
    implemented in this work. Around each grid point the potential is
    approximated to first order. The propagator for a constant force field
    is known analytically. Starting from all possible pairs of initial
    conditions $(q_i,p_j), i,j=1,\ldots N$, the trajectories are deposited
    in momentum space, matched to the momentum grid by 
    $\Lambda(p,\check{p}_j)$. The reconstruction of the wave function in 
    coordinate space is done by inverse discrete Fourier transform.
  }
  \label{fig:sketch_QPIC}
\end{figure}

Aiming at the description of a many-particle system, the numerical effort
for a full quantum calculation increases drastically.
If the system is described in terms of (correctly symmetrized) product states,
the main numerical problem is the exponential growth of the Hilbert 
space dimension with the number of particles.
For some systems, the full correlations encoded in those
states are of minor importance as particular aspects of interest may be described
on a lower level of complexity.
Therefore,  
much interest has been devoted to finding a suitable semiclassical
approximation for the time evolution operator.
%
Based on a path integral description of quantum mechanics~\cite{Fe48,FH65}, 
the Suzuki-Trotter decomposition~\cite{Tr59,Su76} of $U(t-t_0)$ opens the road 
toward a class of semiclassical approximations.
Instead of directly propagating the quantum state as a whole using $U(t-t_0)$, 
we consider the propagation of several individual paths
from $q_0(t_0)$ to $q(t)$ (virtual particle trajectories).
The corresponding propagator is given by
\begin{equation}
  \Pi(q,t;q_0,t_0) \sim \sum\limits_{\substack{\mathrm{v. paths}\\q_0\curvearrowright q}}
  C \exp\left(\imag S(q,t;q_0,t_0)/\hbar\right)\,,
  \label{eq:prop_allg}
\end{equation}
with some normalization factor $C$ and the action
\begin{equation}
 S(q,t;q_0,t_0) = \int\limits_{t_0}^t \D t' p(t')\dot{q}(t') - 
 H\left[p(t'),q(t')\right]\,,
\end{equation}
evaluated along each virtual path.
$H(p,q)$ is the classical Hamilton function of the system.
The only restriction for the virtual paths are fixed starting and end 
points, $q_0(t_0)$ and $q(t)$, apart from which they are completely arbitrary.
Specifically, they do not follow the Hamilton equations of motion, 
which only hold for a subset of them, namely the classical paths
(left panel of Fig.~\ref{fig:sketch_QPIC}). 

In a numerical implementation (middle panel of Fig.~\ref{fig:sketch_QPIC}),
the quantum state is represented on a discrete coordinate grid $\check{q}_i$ 
for any time grid point, while the coordinates of the virtual trajectories are 
treated as continuous variables. 
Reconstructing the quantum state at time $t$ involves the integration
over the propagators between all possible combinations of $q_0$ and $q$.
While we choose the $q_0$ from the set of discrete grid points, the 
end points $q$ may also be off the grid.
For the deposition~\cite{HE88} of a trajectory to the $\check{q}_i$ 
grid, a suitable shape function $\Lambda(\check{q}_i,q)$, 
e.g., a Gaussian or simply a delta peak at the nearest grid point, is 
necessary,
\begin{equation}
  \psi(\check{q}_i,t) = \int \D q_0 \int \D q \;\Lambda(\check{q}_i,q) 
 \Pi(q,t;q_0,t_0) \psi(q_0,t_0)\,.
 \label{eq:prop_grid}
\end{equation}
As the contributions to $\psi(\check{q}_i,t)$ from the different paths are
complex valued, their superposition includes interference effects 
in the reconstructed quantum state.
The major difference between existing semiclassical propagator methods 
concerns the numerical implementation of~(\ref{eq:prop_grid}). 
The standard line of argumentation in the literature is that the main 
contributions to (\ref{eq:prop_allg}) are those for which $S$ is extremal
 -- these are just the classically realized trajectories -- and the paths in their vicinity.
This requirement is deduced from stationary phase integration.
Then, the action is expanded to second order around the extremal
trajectories giving the well known WKB~\cite{We26,Kr26,Br26} result~\cite{Sc81,Gu90}
%
\begin{equation}
  \Pi^{\mathrm{WKB}}(q,t;q_0,t_0) =
  \sum\limits_{\substack{\mathrm{c. paths}\\q_0\curvearrowright q}}
  \sqrt{\frac{\imag}{2\pi\hbar}
    \frac{\partial^2 S_c}{\partial q_0 \partial q}}
    \exp\left(\frac{\imag}{\hbar} S_c(q,t;q_0,t_0)\right)\,,
\end{equation}
where $S_c(q,t;q_0,t_0)$ is the action along the extremal trajectories from 
$q_0(t_0)$ to $q(t)$ and the phase of $\sqrt{\imag}$ is fixed to 
$\pi/4$.
The sum accounts for the fact that specifying $q_0$ and $q$ does not uniquely
determine a classical trajectory.
In addition to the phase factor from the classical action, the phase of the
propagator is determined by the sign of $\partial^2S_c/(\partial q_0\partial q)$.
Using $p_0 = -\partial S_c/\partial q_0$, the Morse theory~\cite{Mi63,Mo73}
allows for separating the determinant of the monodromy matrix 
$\partial q/\partial p_0$ from its sign.
Relating the number of focal points (at which $\partial q/\partial p_0$
vanishes) along a trajectory to the number of negative eigenvalues of 
$\partial^2S_c/(\partial q\partial q_0)$, the whole phase information can be 
encoded in the Morse index $\nu$.
For each focal point, $\nu$ is increased by one, and the square root is 
taken from the absolute value of $\partial q/\partial p_0$.
Then we obtain the Van Vleck Gutzwiller propagator~\cite{Va28,Gu67}
\begin{equation}
  \Pi^{\mathrm{VG}}(q,t;q_0,t_0) = \frac{1}{\sqrt{2\pi\imag\hbar}}
  \sum\limits_{\substack{\mathrm{c. paths}\\q_0\curvearrowright q}}
  \left|\frac{\partial q}{\partial p_0}\right|^{-1/2}
  \E^{\imag S_c(q,t;q_0,t_0)/\hbar} \E^{-\imag\pi \nu/2}\,.
  \label{VV_prop}
\end{equation}
In view of a numerical evaluation, this formulation has several shortcomings.
First, the so called root search problem consists in finding all initial momenta
$p_0$ for which trajectories starting in $q_0$ end up in $q$.
For complex systems, this boundary value problem is much more 
demanding than the solution of an initial value problem.
Second, at the focal points 
the expression (\ref{VV_prop}) diverges and the expansion of $S$ has to be taken 
to higher order.
At those points it is particularly difficult to keep track of the correct branch
of the square root of $\partial q/\partial p_0$ and thus determining the
Morse index.
Both problems are circumvented in contemporary applications of the semiclassical
propagator by expressing (\ref{VV_prop}) in an initial value representation 
(SCIVR)~\cite{Mi01,TW04,Ka94b}.
Inserting (\ref{VV_prop}) into (\ref{eq:prop_grid}), this is achieved by a 
change of variables, 
\begin{equation}
  \int \D q_0 \int \D q \sum\limits_{\substack{\mathrm{c. paths}\\q_0\curvearrowright q}}
  \rightarrow \int \D q_0 \int \D p_0 
  \left|\frac{\partial q}{\partial p_0}\right|\,,
\end{equation}
circumventing the root search, and the monodromy matrix appears
in the numerator,
\begin{equation}
\begin{split}
  \psi^{\mathrm{SCIVR}}(\check{q}_i,t) =  \frac{1}{\sqrt{2\pi\imag\hbar}}
  \int \D q_0 \int \D p_0 \;
   & \Lambda(\check{q}_i,q(q_0,p_0))
  \left|\frac{\partial q(q_0,p_0)}{\partial p_0}\right|^{1/2}\\[0.2cm]
     & \times\E^{\imag S(t;q_0,p_0,t_0)/\hbar} \E^{-\imag\pi \nu/2}
 \psi(q_0,t_0)\,.
\end{split}
\end{equation}
Here, the final coordinate $q(q_0,p_0)$ is a
deterministic function of the initial values $q_0$, $p_0$.
Instead of using a coordinate space basis, similar methods have been
formulated in momentum or coherent state representation~\cite{Ka94b,SWM98,WMM01},
leading to the Herman-Kluk propagator~\cite{HK84,DE06}.
For all these methods the problem of the oscillatory integrand still persists.
To improve the convergence properties of the MC integrals, various 
integral filtering techniques have been proposed in the literature~\cite{Fi86,WMM01,MM87,WM96}.
Thereby, the basic idea is to filter out the high frequency oscillations of the 
integrand which contribute only little to the integral but are the main obstacle
for an efficient MC evaluation.

In this work, we follow a recently proposed~\cite{DDD05,TDD04}, slightly 
different route (cf. right panel of Fig.~\ref{fig:sketch_QPIC}).
Discretizing the potential on the coordinate grid $\check{q}_i$, we locally 
approximate the potential to first order and use this approximation for
$\check{q}_{i}-\Delta\check{q}/2 < q <\check{q}_{i}+\Delta\check{q}/2$.
A justified question then is, to which extend the concatenated potential will 
be able to fully describe quantum effects and reproduce the 
results for the original continuous potential.
The propagator for a trajectory in a linear potential is known
analytically~\cite{GS98},
 \begin{equation}
\begin{split}
   \Pi^{\mathrm{lin}}(q,t;q_0,t_0) = & \sqrt{\frac{m}{2\pi\imag\hbar(t-t_0)}}\\& \times
     \exp\left[\frac{\imag}{\hbar}\left(\frac{m(q-q_0)^2}{2(t-t_0)} - 
         \frac{s}{2}(q+q_0)(t-t_0) - \frac{s^2(t-t_0)^3}{24m}\right)\right]\,,
 \label{eq:prop_lin}
\end{split}
 \end{equation}
where $s$ is the slope of the potential.
In contrast to the WKB formulation, here tracking the sign of the monodromy 
matrix is not necessary, but the maximum possible time step is severely reduced.
This limit is set by the validity of the local potential approximation,
i.e., by the potential variation and grid spacing.
The time step has to be chosen such that even the fastest
particles stay within their initial grid cells intermediately.
An advantage of the constant force field approximation is the bijection 
between final positions and initial momenta which allows for a straight 
forward formulation of the algorithm in terms of an initial value representation.
The range of initial momenta, which are required by Fourier completeness is
given by the coordinate grid spacing as 
$|p_0|\le \pi\hbar/\Delta\check{q}$.
A final trick consists in depositing the contributions from all virtual
particles not directly to the coordinate grid.
Instead, they are gathered in momentum space and 
the reconstructed wave function in coordinate space is obtained 
using inverse discrete Fourier transform,
\begin{equation}
  \psi(\check{q}_i,t) = 
  \sum\limits_{j=1}^{N} \frac{\Delta \check{p}}{\sqrt{2\pi\hbar}}
  \;\exp\left(\frac{2\pi\imag \check{q}_i \check{p}_j}{\hbar N}\right)
   \psi(\check{p}_j,t)\,.
 \end{equation}
This mixed representation has proven less noisy than a
direct deposition in coordinate space~\cite{Ka94b}.
In analogy to (\ref{eq:prop_grid}), we use the shape function 
$\Lambda(\check{p}_j,p)$ for the deposition on the momentum grid
\begin{equation}
  \psi(\check{p}_j,t) = \int \D q_0 \int \D p \;\Lambda(\check{p}_j,p) 
 \Pi^{\mathrm{lin}}(p,t;q_0,t_0) \psi(q_0,t_0)\,,
\end{equation}
where $\Pi^{\mathrm{lin}}(p,t;q_0,t_0)$ is the Fourier transform
of (\ref{eq:prop_lin}),
\begin{equation}
  \Pi^{\mathrm{lin}}(p,t;q_0,t_0) = \frac{1}{\sqrt{2\pi\hbar}} \int 
  \D q \; \Pi^{\mathrm{lin}}(q,t;q_0,t_0) \E^{-\imag pq/\hbar}\,.
  \label{eq:lin_prop}
\end{equation}
Instead of an explicit evaluation of the $q$-integral in (\ref{eq:lin_prop}), we profit from the
knowledge about the trajectories,
\begin{equation}
  q(t) = -\frac{s}{2m}(t-t_0)^2 + \frac{p_0}{m} (t-t_0) + q_0\,,
\label{eq:traj_q}
\end{equation}
to change the integration variable from $q$ to $p_0$.
Finally, this allows us to propagate the wave function by one time step,
$\Delta t= t-t_0$, using an initial value representation,
\begin{equation}
\begin{split}
  \psi^{\mathrm {lin}}(\check{q}_i,t) = 
  \sum\limits_{j=1}^{N} \frac{\Delta \check{p}}{2\pi\hbar}
  \exp\left(\frac{2\pi\imag \check{q}_i \check{p}_j}{\hbar N}\right)
  \int \D p\;\Lambda(\check{p}_j,p) 
  \sqrt{\frac{1}{2\pi\imag\hbar}\frac{\Delta t}{m}}
  \int \D q_0 \int \D p_0\; \psi(q_0,t_0)\\
  \times \exp\left[\frac{\imag}{\hbar}\left(\frac{m[q(q_0,p_0)-q_0]^2}{2\Delta t} - 
      \frac{s\Delta t}{2}[q(q_0,p_0)+q_0] - \frac{s^2(\Delta t)^3}{24m} -pq(q_0,p_0)\right)\right]\,,
\end{split}
\end{equation}
where $q(q_0,p_0)$ is given in (\ref{eq:traj_q}).

\subsection{Wigner-Moyal approach}
\label{sect:Wig}
Dating back to Wigner~\cite{Wi32}, the idea of describing 
quantum mechanics without a complicated operator algebra,
but by equations for commuting variables, has been very appealing.
Then, introducing a phase space distribution function and replacing operator
expressions according to the corresponding rules of association, quantum 
expectation values may be calculated by simple integration over commuting 
variables.
One of the most common~\cite{Le95} distribution functions
is the Wigner function,
\begin{equation}
  W(q,p,t) = \frac{1}{2\pi} \int \E^{-\imag\eta p} 
  \psi^{\star}(q-\eta\hbar/2,t) \psi(q+\eta\hbar/2,t)\, \D\eta\,,
  \label{eq:Wigner}
\end{equation}
together with the corresponding Weyl rule of association~\cite{We31}.
%
Within this rule of association, operator expressions ordered as 
$\E^{\imag\xi\hat{q}+\imag\eta\hat{p}}$ are replaced by their 
scalar variables, e.g., 
$\E^{\imag\xi\hat{q}+\imag\eta\hat{p}} \leftrightarrow 
\E^{\imag\xi q+\imag\eta p}$, with $\xi,\eta\in \mathbb{C}$.
Starting from the von Neumann equation for the time evolution of the 
density matrix, we determine the evolution equation for the Wigner 
function~\cite{Fi96c,FBFG08,Mo49b}
as
\begin{equation}
\frac{\partial W}{\partial t}+ \frac{p}{m} \frac{\partial W}{\partial q} 
+ F(q) \frac{\partial W}{\partial p} = \int\limits_{-\infty }^\infty 
 \D s\;W(q, p-s, t) \omega(s,q)\,,
\label{EOM_Wigner}
\end{equation}
where $F(q)=-\frac{dV(q)}{dq}$ is the classical force, and 
\begin{equation}
  \omega(s,q) = \frac{2}{\pi\hbar^2}
  \int \D q'\; V(q-q') \sin \left( \frac{2 sq'}{\hbar}\right) + F(q) \frac{d\delta(s)}{ds}\,.
  \label{eq:omega}
\end{equation}
In the classical limit, the right hand side of (\ref{EOM_Wigner}) vanishes, leaving
us with the Liouville equation for the phase space density.
Then the  dynamics can be expressed in terms of the classical 
propagator
\begin{equation}
  \Pi^W(q,p,t;q_0,p_0,t_0) =\delta [q-\bar{q}(t;p_0,q_0,t_0)]
  \delta [p-\bar{p}(t;p_0,q_0,t_0)] 
  \,.
\end{equation}
Here $\bar{p}$ and $\bar{q}$ are the momentum and coordinate of a trajectory which
evolves according to the Hamilton equations of motion subject to the initial
conditions $\bar{p}(t_0)=p_0$ and $\bar{q}(t_0)=q_0$.
Using $\Pi^W$, we may rewrite (\ref{EOM_Wigner}) in form of an 
integral equation~\cite{FMK95,Fi96c,FBFG08},
\begin{equation}
\begin{split}
W(q,p,t) &= \int  \D p_0\,  \D q_0 \;\Pi^W(q,p,t;q_0,p_0,t_0)
W_0(q_0,p_0,t_0)  \\&
 +  \int\limits_{t_0}^t d\tau \int \D p_{\tau} \D q_{\tau}\;
\Pi^W (q,p,t; q_{\tau},p_{\tau},\tau) 
\int\limits_{-\infty }^\infty \D s\; W(q_{\tau}, p_{\tau}-s,\tau)
 \omega (s,q_{\tau})  \label{s6}\,,
\end{split}
\end{equation}
and solve it by iteration. 
To lowest order, we only keep the first line in (\ref{s6}) and neglect the
second integral completely.
This means, we propagate classical trajectories $(\bar{q},\bar{p})$ in time,
after sampling their initial conditions $p_0$ and $q_0$ from the initial
Wigner function $W_0(q_0,p_0,t_0)$ at time $t_0$ using a MC procedure.
Assembling those contributions at the next time grid point $t=t_0+\Delta t$,
we obtain the lowest order approximation $W^{(1)}(q,p,t)$.

For the second order approximation, $W^{(2)}(q,p,t)$, we expand the Wigner
function in the last integral in~(\ref{s6}) consistently to lowest order,
\begin{equation}
\begin{split}
  W( q_{\tau}, p_{\tau}-s,\tau)\approx &  W^{(1)}( q_{\tau}, p_{\tau}-s,\tau)  
  \\ =  &  \int  \D p_0\,  \D q_0 \;
  \Pi^W(q_\tau, p_{\tau}-s,\tau;q_0,p_0,t_0) W_0(q_0,p_0,t_0)\,.
  \label{eq:second_order}
\end{split}
\end{equation}

\begin{figure}
  \centering
  \includegraphics[width=0.5\linewidth,clip]{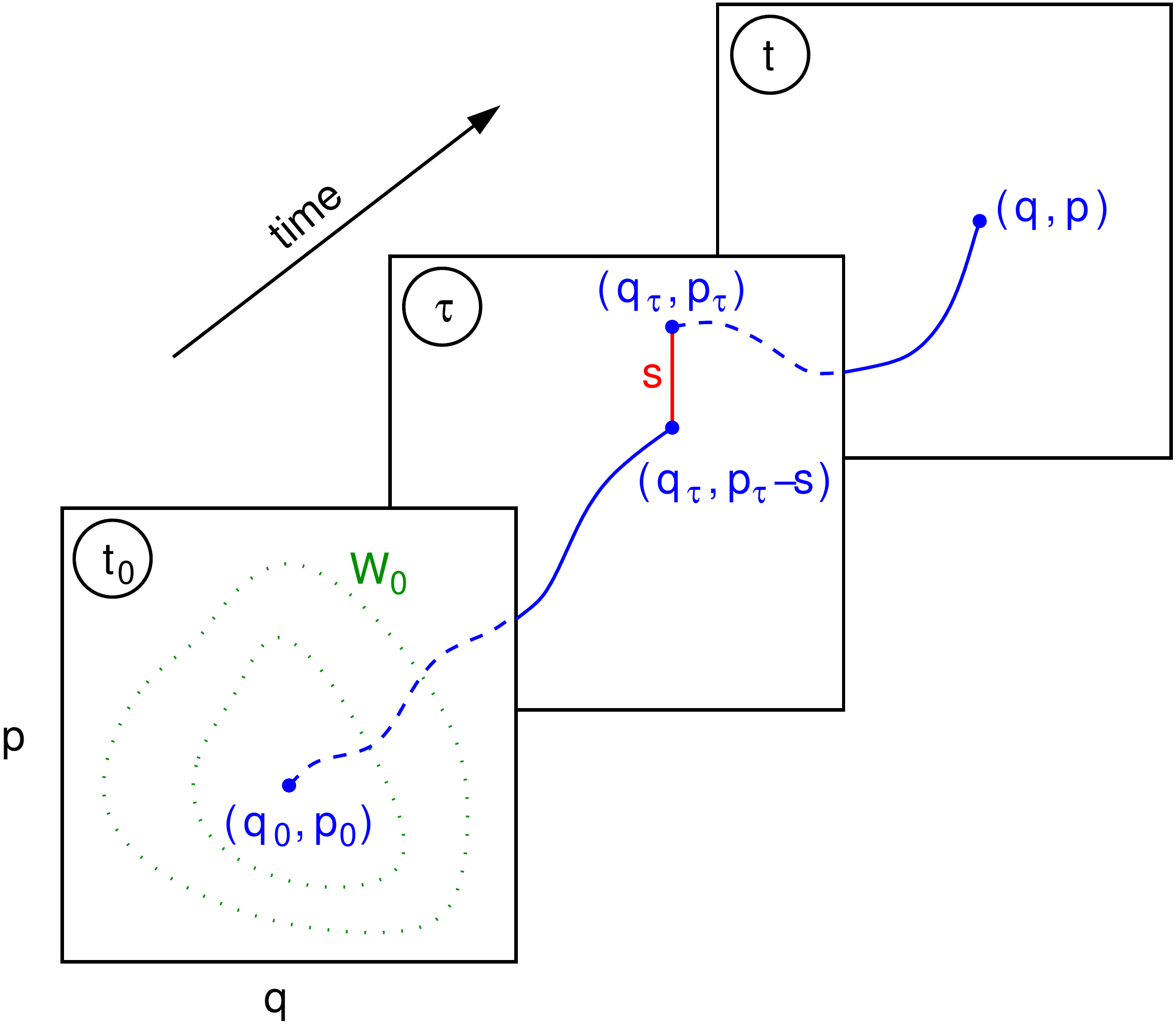}
  \caption{Cartoon of propagating one trajectory in 
    presence of momentum jumps.
    At time $t_0$, the initial conditions for the trajectory $(q_0,p_0)$ are
    sampled from $W_0(q_0,p_0,t_0)$ by a MC procedure.
    Then up to time $\tau\in[t_0,t]$ this trajectory is propagated to 
    $(q_\tau, p_{\tau}-s)$ according to the classical equations of motion.
    There, an instantaneous change in momentum by $s$ occurs, leaving the 
    intermediate position $q_\tau$ fixed. 
    From the new phase space point $(q_{\tau},p_{\tau})$ the trajectory
    evolves again classically up to $t$, following the propagator 
    $\Pi^W (q,p,t; q_{\tau},p_{\tau},\tau)$.
    When summing up the contributions of all trajectories, each one has to be
    weighted by the function $\omega(s,q_\tau)$, accounting for the momentum
    jump $s$ and the potential at the intermediate position $q_{\tau}$.
  }
  \label{fig:scetch_wig}
\end{figure}
This allows for evaluating $W^{(2)}$ also by means of
trajectory methods. 
A sketch of the basic idea behind the second order approximation is given in 
Fig.~\ref{fig:scetch_wig}, together with a detailed description in the 
caption.
Including momentum jumps in the evolution of $W^{(2)}$ we take into 
account that continuous phase space trajectories are guaranteed
only for classical or almost classical systems~\cite{Le95}. 
Depending on the momentum jump $s$ and the potential $V(q_{\tau}-q')$, 
the weighting factor $\omega(s,q_\tau)$ may become negative.
Therefore, sign changes of the Wigner function at some phase space points
are included in the second order approximation, whereas they are absent
in the first order approximation due to strictly positive trajectory weights.

Using the improved estimate  $W^{(2)}$ in (\ref{eq:second_order}), the construction
of higher order terms is straight forward.
Conceptually, higher order terms correspond to the inclusion of 
several momentum jumps within one time step.
As the fundamental aspects (discontinuous trajectories, possibility of 
negative trajectory weights) are already included in $W^{(2)}$,
we restrict ourselves to the two lowest orders of the approximation.
A detailed investigation of the inclusion of higher order terms is 
beyond the scope of this work (for details see~\cite{FMK95,Fi96c}).

Having $W(q,p,t)$ at hand, it is not difficult to calculate expectation values
of all kinds of operators from it. 
For any operator expression which is ordered as 
$\E^{\imag \xi\hat{q} + \imag\eta\hat{p}}$, we follow the Weyl rule of association
and replace each operator by its corresponding scalar function.
After multiplication with the Wigner function we integrate over the whole
phase space and obtain the desired expectation value.

\subsection{Tomographic approach}
\label{sect:TOM}

\begin{figure}
  \centering
  \includegraphics[width=\linewidth,clip]{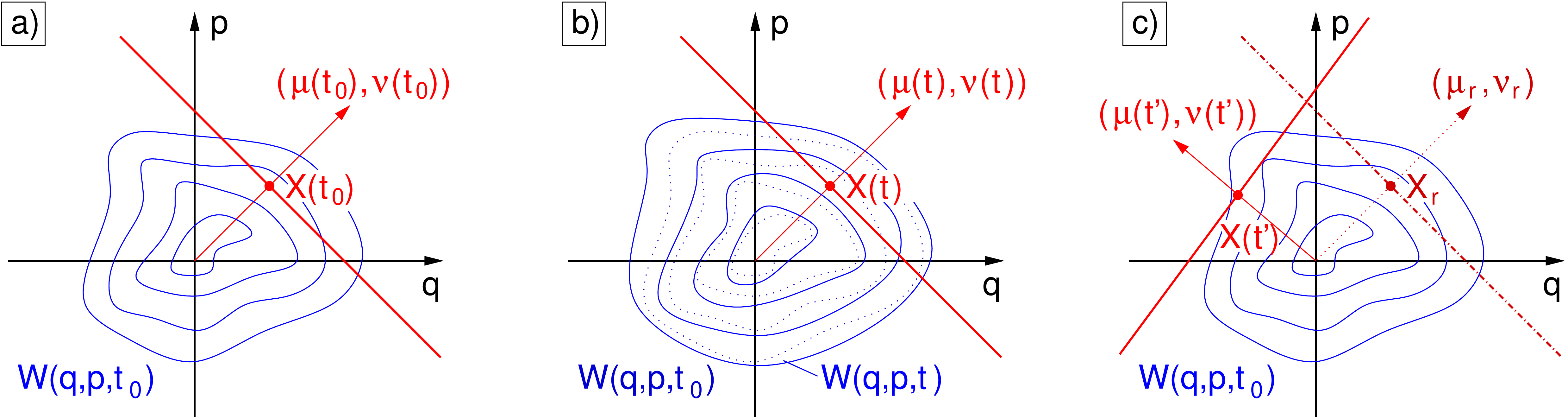}
  \caption{Cartoon of the quantum tomogram and
    its time evolution. Panel (a): The tomogram 
    $\tilde{w}(X(t_0),\mu(t_0),\nu(t_0),t_0)$ is the integral over 
    $W(q,p,t_0)$ along the thick red line. Panel (b): Extraction of the 
    tomogram in the same reference frame at time~$t$ by evolving the Wigner
    function and keeping $(\mu_r,\nu_r)=[\mu(t),\nu(t)]=[\mu(t_0),\nu(t_0)]$ fixed. 
    Panel (c): Time evolution of 
    the tomogram by the method of characteristics. Keeping the Wigner function
    $W(q,p,t_0)$ fixed, $(\mu_r,\nu_r)=[\mu(t),\nu(t)]$ is propagated backward in 
    time to $(\mu(t'),\nu(t'))$ using (\ref{eq:TOM_prop_H}) for different local
    approximations. Summing all those contributions gives the tomogram at
    $t$ in the desired reference frame $(\mu_r,\nu_r)$.}
  \label{fig:TOM_scetch}
\end{figure}

In view of a probabilistic interpretation of quantum mechanics, the 
Wigner function has the shortcoming of possible negative values, which 
prevents its interpretation as a probability density.
Furthermore, any interpretation of a joint probability, that simultaneously
determines coordinate and momentum for a quantum system, violates the
Heisenberg uncertainty relation and is therefore misleading.
Meaningful results for probabilities and expectation values,
are integrals over the Wigner function in extended phase space regions,
e.g., minimum uncertainty Gaussians, or along phase space contours.
Such a contour integration is used in the 
tomographic representation of quantum mechanics
proposed some years ago~\cite{MMT96,MMT97}.
The so called quantum tomogram~\cite{ALM04,AL03}, 
\begin{equation}
\label{eq:tom}
  \tilde{w}(X,\mu,\nu,t) = \int \frac{\D k\, \D q\, \D p }{2\pi}
    W(q,p,t) \E^{-\imag k(X - \mu q - \nu p)}\,,
\end{equation}
relates to the Wigner function by a class of Radon transformations~\cite{De83}
which are characterized by $\mu$ and $\nu$.
A simple, intuitive interpretation of the quantum tomogram is shown 
in Fig.~\ref{fig:TOM_scetch}(a).
The vector $(\mu,\nu)$ fixes a direction, with $X$ the distance from the origin.
Then the tomogram $\tilde{w}(X,\mu,\nu)$ is just the integral over
the Wigner function along a straight line perpendicular to $(\mu,\nu)$
which passes through $X$.
Choosing $(\mu,\nu)$ appropriately, we may continuously change between 
coordinate and momentum representation.
Describing a quantum system in terms of the tomogram is completely 
equivalent to the wave function or Wigner function representation.
Its relation to the Wigner function is obvious due to its definition,
and we obtain the Wigner function from the tomogram by the inverse map
\begin{equation}
  W(q,p,t)  = \int \frac{\D X\, \D\mu\, \D\nu }
  {(2\pi)^{2}}
  \tilde{w}(X, \mu, \nu,t) \E^{\imag (X - \mu q - \nu p)}\,.
\end{equation}
The relation to the wave function formalism is a little more involved.
We may extract the probability densities in any rotated reference frame
(in particular coordinate and momentum space) from the tomogram, but not 
the wave function itself.
The phase information inherent to the wave function is distributed over 
all reference frames ($\mu,\nu$).
Each fixed choice of $(\mu,\nu)$ gives only a density information, e.g., 
the coordinate representation $|\psi(q)|^2$ is equal to 
$\tilde{w}(X, \mu=1, \nu=0)$.
Keeping this aspect in mind as well as the larger required set of variables,
one might ask at this point what the profit of this method shall be.
Considering the dynamics of a quantum system, the strict positivity of the
tomogram is its main numerical advantage.
This circumvents the dynamical sign problem encountered in the
Wigner function and semiclassical propagator approaches.

The time evolution of the tomogram can be thought of in two equivalent ways.
For any reference frame $(\mu(t_0),\nu(t_0))$, e.g., the coordinate representation 
$(1,0)$, the tomogram can be calculated for all times by keeping
$(\mu,\nu)$ fixed and evolving the Wigner function in 
time~[Fig.~\ref{fig:TOM_scetch}(b)].
This is exactly the interpretation we used in Sect.~\ref{sect:Wig} to extract 
$|\psi(q,t)|^2$ from the Wigner function.
Alternatively, we can exploit the evolution equation of the 
tomogram~\cite{FSLBFFF08,AM04},
\begin{equation}
  \frac{\partial \tilde{w}}{\partial t} -
  \frac{\mu}{m}\frac{\partial \tilde{w}}{\partial \nu} -
  \frac{\imag}{\hbar}  \left[V \left(
      - \frac{\partial }{\partial \mu} \frac{1}{\partial / \partial X}
      - \frac{ \imag\hbar\nu}{2} \frac{\partial }{\partial X} \right) -
    V\left( - \frac{\partial }{\partial \mu} \frac{1}{\partial /
        \partial X} + \frac{ \imag\hbar\nu}{2} \frac{
        \partial }{\partial X} \right) \right ] \tilde{w} = 0\,,
  \label{eq:EOM_TOM}
\end{equation}
which can be derived from the von Neumann equation.
%
In view of the definition of $\tilde{w}$ in (\ref{eq:tom}), 
the `anti-derivative' $\partial_X^{-1}\tilde{w}$ just 
multiplies $W(q,p)$ by $\imag/k$.
Therefore, the term $\partial_\mu \partial_X^{-1}\tilde{w}$ is well defined.
%
A solution of (\ref{eq:EOM_TOM}) for harmonic potentials
$V(q) = \frac{1}{2}m\omega_0^2(q-q_c)^2$ can be given explicitly. 
In analogy to the continuity equation for the tomogram,
\begin{equation}
  \label{eq:CEQ_TOM}
  \frac{d\tilde{w}}{dt} =  \frac{\partial\tilde{w}}{\partial t} + 
  \frac{\partial\tilde{w}}{\partial X} \dot{X} +
  \frac{\partial\tilde{w}}{\partial \mu} \dot{\mu} + 
  \frac{\partial\tilde{w}}{\partial \nu} \dot{\nu} = 0\,,
\end{equation}
we collect the terms in (\ref{eq:EOM_TOM}) and find the correspondences
\begin{equation}
  \dot{X} = m\omega_0^2 q_c\nu\quad ,\quad \dot{\mu}= m\omega_0^2 \nu \,
  \quad,\quad\ \dot{\nu}=-\mu/m\,.
  \label{eq:EOM_TOM_traj}
\end{equation}
This set of linear differential equations defines trajectories in 
$(X,\mu,\nu)$ space characterizing the reference frames in
Fig.~\ref{fig:TOM_scetch}.
Solving the system (\ref{eq:EOM_TOM_traj}), we get the 
propagator from $(X_0,\mu_0,\nu_0,t_0)$ to $(X,\mu,\nu,t)$  as
\begin{align}
  \Pi^{\mathrm{T}}_{\omega_0,q_c}(X,\mu,\nu,t;X_0,\mu_0,\nu_0,t_0) =
  \delta \left[ \nu + \frac{\mu_0}{m\omega_0}\sin[\omega_0 (t\!-\!t_0)] - 
    \nu_0\cos[\omega_0 (t\!-\!t_0)]\right]& \nonumber\\
  \times \delta\Big[X-X_0-\mu_0 q_c\big(\cos[\omega_0(t\!-\!t_0)]-1\big) - 
  \nu_0 m\omega_0 q_c\sin[\omega_0(t\!-\!t_0)]\Big]& \nonumber \\
   \times  \delta \Big[\mu - \mu_0 \cos[\omega_0 (t\!-\!t_0)] + 
    \nu_0 m\omega_0\sin[\omega_0 (t\!-\!t_0)]\Big]&\,. 
  \label{eq:TOM_prop_H}
\end{align}
The uniqueness of the $(X(t),\mu(t),\nu(t))$ trajectories is due to the
$q$-independency of the coefficients in the set of equations (\ref{eq:EOM_TOM_traj}).

Evaluating (\ref{eq:EOM_TOM}) for arbitrary potentials (beyond the free particle 
or harmonic case), higher order derivatives of $\tilde{w}$ appear. 
This spoils the identification with the continuity equation (\ref{eq:CEQ_TOM})
and the application of the method of characteristics (\ref{eq:EOM_TOM_traj}).
A possible way out is the local expansion of the potential to second order.
Then (\ref{eq:EOM_TOM})-(\ref{eq:TOM_prop_H}) are valid locally for each $q$
with appropriate parameters $\omega_0(q)$ and $q_c(q)$ which are determined
from the harmonic expansion around $q$.
Since now several propagators (\ref{eq:TOM_prop_H}) exist, the 
$(X,\mu,\nu)$ trajectories are not unique anymore, and the tomogram at
the new time grid point is given by
\begin{equation}
\begin{split}
  \tilde{w}(X,\mu,\nu,t) = &\int \D X_0 \int \D \mu_0 \int \D \nu_0 \int \D q\;
   \\ & \Pi^{\mathrm{T}}_{\omega_0(q),q_c(q)}(X,\mu,\nu,t;X_0,\mu_0,\nu_0,t_0)
   \tilde{w}(X_0,\mu_0,\nu_0,t_0)\,.
\label{eq:TOM_total}
\end{split}
\end{equation}

Usually, we are not interested in the full $\tilde{w}(X,\mu,\nu,t)$ but
only in a particular reference frame $(\mu_r,\nu_r)$, 
e.g., the coordinate representation.
Having the method of characteristics in mind, we search for those
trajectories $(X(t_0),\mu(t_0),\nu(t_0))$ that give contributions to 
$(X(t),\mu(t),\nu(t)) = (X_r,\mu_r,\nu_r)$.
To this end, we start from the known 
$\tilde{w}(X_r,\mu_r,\nu_r,t_0)$ and evolve the trajectories
backward in time to $t'= t_0-(t-t_0)$ according to (\ref{eq:TOM_prop_H})
using different coordinates $q$ [cf. Fig.~\ref{fig:TOM_scetch}(c)].
The desired $\tilde{w}(X_r,\mu_r,\nu_r,t)$ is then
the sum over all such tomograms $\tilde{w}(X_q(t'),\mu_q(t'),\nu_q(t'),t_0)$,
where the index $q$ reflects the dependency on the 
parameters $\omega_0(q)$ and $q_c(q)$ in the propagator 
$\Pi^{\mathrm{T}}_{\omega_0(q),q_c(q)}$.

An alternative approach which relates the time evolution of the quantum 
tomogram to the theory of Markov processes~\cite{Ga90} has been 
considered recently~\cite{FSLBFFF08}.
Interpreting the propagator $\Pi^{\mathrm{T}}$ as a transition probability
for a Markov random process, its time evolution is governed by the Chapman-Kolmogorov
equation~\cite{Sv07}
\begin{equation}
  \Pi^{\mathrm{T}}(z,t;z_0,t_0) = \int  \D z_\tau \; \Pi^{\mathrm{T}}(z,t;z_\tau,\tau)
  \Pi^{\mathrm{T}}(z_\tau,\tau;z_0,t_0) \,,
\end{equation}
where $t_0<\tau<t$ and we introduced the abbreviation 
$z = (z^1,z^2, z^3):=(X,\mu,\nu)$.
Due to the positivity of the tomogram and its normalization, the 
requirements
\begin{equation}
  \Pi^{\mathrm{T}}(z,t;z_0,t_0)\geq 0 \quad,  \quad \int  \D z_0\;  \Pi^{\mathrm{T}}(z,t;z_0,t_0) = 1
\end{equation}
for a Markov process are fulfilled. Furthermore, the locality in time 
of the Hamiltonian guarantees that no memory effects are present. 
The dynamics of diffusive Markov processes may be equivalently described by 
two partial differential equations, the first and the second Kolmogorov
equation~\cite{Ko31},
\begin{eqnarray}
\frac{\partial \Pi^{\mathrm{T}}}{\partial t_0}+\sum_{i=1}^3 a^i(z_0,t_0)
\frac{\partial \Pi^{\mathrm{T}}}{\partial z_{0}^i} + \frac{1}{2} \sum_{i=1}^3
\sum_{j=1}^3 b^{ij}(z_0,t_0)\frac{\partial ^2 \Pi^{\mathrm{T}}}{\partial z^i_0
\partial z^j_0 } = 0\,,
\\
\frac{\partial \Pi^{\mathrm{T}}}{\partial t}+\sum_{i=1}^3 \frac{\partial
}{\partial z^i} \left (a(z,t) \Pi^{\mathrm{T}} \right ) -
\frac{1}{2}\sum_{i=1}^3 \sum_{j=1}^3 \frac{\partial ^2 }{\partial
z^i \partial z^j } \left( b(z,t) \Pi^{\mathrm{T}} \right)  = 0\,,
\label{eq:TOM_Kolm}
\end{eqnarray}
in which the drift and diffusion coefficients are defined as
\begin{eqnarray}
a^i(z_\tau,\tau) & = & \lim_{\tau'\to\tau} \frac{1}{\tau'- \tau} \int
\D \tilde{z}_{\tau'}\;
\Pi^{\mathrm{T}} (\tilde{z}_{\tau'},\tau';z_\tau,\tau)(\tilde{z}^i_{\tau'}-z^i_\tau) 
\label{eq:TOM_drift}\,,\\
b^{ij}(z_\tau,\tau) &= & \lim_{\tau'\to\tau} \frac{1}{\tau'-\tau} \int
\D \tilde{z}_{\tau'}\;
\Pi^{\mathrm{T}} (\tilde{z}_{\tau'},\tau';z_\tau,\tau) (\tilde{z}^i_{\tau'}-z^i_\tau)
(\tilde{z}^j_{\tau'}-z^j_{\tau})\,.
\label{eq:TOM_diff}
\end{eqnarray}
In the limit $\tau'\to\tau$ we can evaluate~(\ref{eq:TOM_drift}) and
(\ref{eq:TOM_diff}) using the harmonic approximation 
$\Pi^{\mathrm{T}}_{\omega_0,q_c}$ from~(\ref{eq:TOM_prop_H}).
As the harmonic propagator depends on $\omega_0(q)$ and $q_c(q)$ of
the local potential expansion, also the drift and diffusion terms are 
$q$ dependent.
For clarity of the notation, we will suppress the additional index 
for the moment.

Instead of solving (\ref{eq:TOM_Kolm}) directly, we consider the 
equivalent system of stochastic integral
equations~\cite{Ga90,Sv07} for the underlying random variables $z$,
\begin{equation}
  z^i(t) = z^i(t_0) +  \int\limits_{t_0}^{t} \D \tau\; \psi(z_\tau,\tau)
  + \sum\limits_{j=1}^{3} \int\limits_{t_0}^{t} \D \tau \;
  g^{ij}(z_\tau,\tau)\xi^j(\tau)\,.
  \label{eq:TOM_SIE}
\end{equation}
Here $\xi^j(\tau)$ are white noise random processes with 
$\langle\xi^j(\tau)\rangle = 0$ and 
$\langle\xi^j(\tau) \xi^k(\tau')\rangle = \delta(\tau-\tau')\delta^{jk}$.
For evaluating the second integral we will refer to 
the Stratonovich definition of a stochastic integral~\cite{Ga90}.
The functions $\psi(z_\tau,\tau)$ and $g^{ij}(z_\tau,\tau)$ relate to 
the drift and diffusion coefficients in~(\ref{eq:TOM_drift}) and
(\ref{eq:TOM_diff}) as
\begin{eqnarray}
  a^i(z_\tau,\tau)   & = & \psi^i(z_\tau,\tau) + \frac{1}{2}\sum\limits_{j=1}^{3}
  \sum\limits_{k=1}^{3} 
  \frac{\partial g^{ik}(z_\tau,\tau)}{\partial z_\tau^j}g^{jk}(z_\tau,\tau)
  \,,\label{eq:TOM_relate_a}\\
  b^{ij}(z_\tau,\tau) & = & \sum\limits_{k=1}^{3} 
  g^{ik}(z_\tau,\tau)g^{jk}(z_\tau,\tau)
  \,.
  \label{eq:TOM_relate_b}
\end{eqnarray}
For an implementation of the derivative term 
in~(\ref{eq:TOM_relate_a}) we profit from the equivalence~\cite{Sv07}
\begin{equation}
 \frac{1}{2}\sum\limits_{j=1}^{3}
  \sum\limits_{k=1}^{3} 
  \frac{\partial g^{ik}(z_\tau,\tau)}{\partial z_\tau^j}g^{jk}(z_\tau,\tau)
  = \lim_{\Delta\tau\to0} \frac{1}{\Delta\tau} \left\langle
    \sum_{k=1}^{3} g^{ik}(
      \bar{z}_\tau
      , \tau)\Delta \xi_{k} (\tau)
\right\rangle\,,
\label{eq:TOM_gdg}
\end{equation}
where $\langle\ldots\rangle$ means the stochastic expectation value over
the normalized Wiener process $\xi_{k}(\tau)$ and $\Delta\xi_{k}(\tau)$
is the corresponding increment of this random process in $\Delta \tau$.
As a consequence of the Stratonovich integration the matrix elements
$g^{ik}$ have to be evaluated at $\bar{z}_\tau = 
z_\tau +\frac{1}{2}\Delta z_\tau$, i.e., shifted by half the increment of $z_\tau$
during $\Delta\tau$.
According to (\ref{eq:TOM_relate_b}) these matrix elements can be calculated
via Cholesky decomposition~\cite{PFTV86} from $b(\bar{z}_\tau,\tau)$.
Note, that the matrix $b$ is only positive semidefinite, requiring some care
in the numerical determination of $g$, e.g, adding $\varepsilon$ to $b$ and
taking the limit $\varepsilon \to 0$.

In summary, the evolution of a $(X,\mu,\nu)$ trajectory is calculated 
iteratively from (\ref{eq:TOM_SIE})-(\ref{eq:TOM_gdg}).
To lowest order, in (\ref{eq:TOM_SIE}) only the deterministic drift
term~(\ref{eq:TOM_drift}) is taken into account, which is evaluated using
the local harmonic propagator (\ref{eq:TOM_prop_H}).
From the resulting increments $z(t)-z(t_0)$ we get a first estimate 
for $b$ (and thus $g$) which we use in the next iteration.
Having access to the deterministic part as well as the diffusion term in 
(\ref{eq:TOM_SIE}) we calculate several trajectories, needed for 
the expectation value in (\ref{eq:TOM_gdg}).
Now an approximation for all terms in (\ref{eq:TOM_SIE}) is available 
and we can finally propagate our $(X,\mu,\nu)$ trajectory.
Alternatively, we can repeat the last step to ensure that the 
increments entering in (\ref{eq:TOM_gdg}) have been calculated using
the full stochastic integral equation.

In addition to the stochastic character of the trajectory propagation, we
have to keep in mind that the drift and diffusion terms depend on
the coordinate at which (\ref{eq:TOM_prop_H}) is evaluated. 
A suitable importance sampling of $q$ is of crucial importance for an
 efficient implementation.

Although we discussed for simplicity only a three-dimensional vector 
$z=(X,\mu,\nu)$, which corresponds to one set of initial conditions, the
generalization to $k$ initial conditions directly carries over.
In any case, including the initial condition for the coordinate representation 
is obligatory as a reconstruction of $q$ out of $z$ is necessary
for intermediate evaluations of~(\ref{eq:TOM_prop_H}).

From the tomogram, general expectation values can be calculated in analogy 
to the Wigner function case~\cite{FSLBFFF08}.
If the desired expectation value involves only position or momentum
operators but not both, this task simplifies to an integration over the
corresponding density.
Let us consider for instance the kinetic energy $\frac{1}{2m}\langle p^2\rangle$.
With $\mu=0$ and $\nu=1$, the integral representation of the 
$\delta$-distribution in (\ref{eq:tom}) shows that $X\equiv p$ and we get 
\begin{equation}
  \frac{1}{2m}\langle p^2\rangle =  \frac{1}{2m} \int \D X\; X^2 
  \tilde{w}(X,\mu=0,\nu=1) = \frac{1}{2m} \int \D p\; p^2 | \psi(p,t)|^2\,.
\end{equation}
%

\section{Numerical Evaluation}
\label{results}
%
%

\begin{figure}
  \centering
  \includegraphics[width=\linewidth,clip]{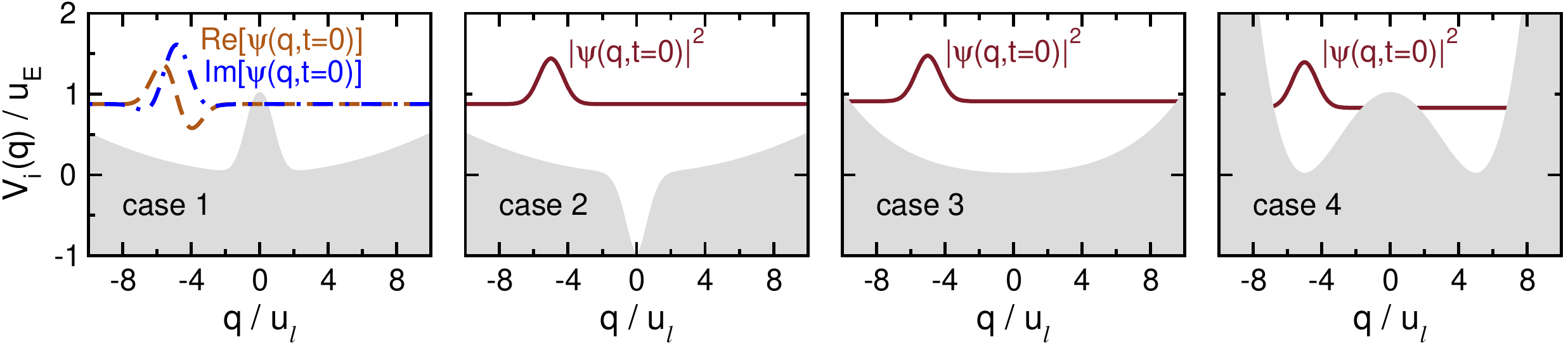}
  \caption{From left to right: Benchmark potentials $V_i(q)$
    as given by~(\ref{pot4}). The initial state in all cases is
    the same. For the first case we show the real and imaginary part
    of the wave function by dashed and dashed-dotted lines.
    In the other panels, the modulus square of the wave function is given. 
    For comparison of the relevant energies, the baseline of 
    the wave function is drawn at the energy of the initial state.}
  \label{fig:Benchmark_pot}
\end{figure}

As test cases for the above methods, we consider one-dimensional
systems described by the Hamilton operators
 $ H_i = \frac{p^2}{2m} + V_i(q)$.
The four potentials
\begin{equation}
\begin{split}
  V_1(q)   = \frac{1}{2}m\omega_0^2q^2 + V_0\exp(-q^2)
  \quad  , &\quad V_2(q)  = \frac{1}{2}m\omega_0^2q^2 - V_0\exp(-q^2)
 \\
  V_3(q) = \frac{1}{2}m\omega_0^2 (q^2 + a_3 q^4)
  \quad  , & \quad
  V_4(q)  = V_0 + \frac{1}{2}m\omega_4^2 (-q^2 + a_4 q^4)\,,
  \label{pot4}
\end{split}
\end{equation}
shown in Fig.~\ref{fig:Benchmark_pot}, each pose another 
numerical difficulty to semiclassical approaches.
It is well known, that semiclassical methods perform well for harmonic
or weakly anharmonic cases, but that the effects of strong 
anharmonicities are hard to capture~\cite{Ma04,SM99}.
Specifically, we chose $V_1(q)$ to study tunneling effects, $V_2(q)$ 
in view of resonances and $V_3(q)$ to investigate the influence of a
nonlinear force term.
Finally $V_4(q)$ combines the challenge of tunneling effects and 
anharmonicities in the potential. 
The inclusion of the shallow harmonic trap in $V_1(q)$ and $V_2(q)$
prevents the particle from escaping to infinity after the scattering event
and restricts the simulation volume to a reasonable size.
For all cases we use the same initial conditions, a Gaussian wave packet
of width $\sigma$, centered at $q_0$, with center momentum $p_0$,
\begin{equation}
  \psi(q,t=0) =  \frac{1}{(2\pi\sigma^2)^{1/4}} 
  \exp\left\{-\frac{1}{4\sigma^2} (q-q_0)^2 + \frac{\mathrm i}{\hbar} p_0 q\right\}\,.
\label{eq:psi0}
\end{equation}
Taking the modulus squared $|\psi(q,t=0)|^2$ of (\ref{eq:psi0}), the
correct normalization to unity is obvious.
Using the initial wave function~(\ref{eq:psi0}) together with (\ref{eq:Wigner}) we get 
the corresponding initial Wigner function
\begin{equation}
  W(q,p,t=0) = \frac{1}{\pi\hbar} \exp\left\{
    -\frac{1}{2\sigma^2}(q-q_0)^2 - \frac{2\sigma^2}{\hbar^2}(p-p_0)^2\right\}\,.
  \label{eq:wig0}
\end{equation}
Comparing the coefficient of $(p-p_0)^2$ in the exponent to the standard
form of a Gaussian,
the standard deviation in momentum space  reads $\sigma_p=\hbar/(2\sigma)$.
Therefore, $\sigma\sigma_p=\hbar/2$, which makes our initial state a minimum 
uncertainty Gaussian wave packet.
Integrating the Wigner function over the momentum (coordinate) variables,
we get the probability density in coordinate (momentum) space,
\begin{equation}
  |\psi(q,t=0)|^2 = \int \D p\; W(q,p,t=0) =  
  \frac{1}{\sqrt{2\pi\sigma^2}}  
  \exp\left\{-\frac{(q-q_0)^2}{2\sigma^2} \right\}\,.
\end{equation}
It is clear, that we could have obtained this expression also directly by
taking the modulus square of (\ref{eq:psi0}).
To obtain the probability in momentum space,
\begin{equation}
  |\psi(p,t=0)|^2 = \int \D q\; W(q,p,t=0) =  
  \sqrt{\frac{2\sigma^2}{\pi\hbar^2}}  
  \exp\left\{ -\frac{2\sigma^2(p-p_0)^2}{\hbar^2} \right\}\,,
\end{equation}
from the wave function, however, first a Fourier transform to momentum
representation is necessary,
\begin{equation}
\begin{split}
  \psi(p,t=0)&  = \frac{1}{\sqrt{2\pi\hbar}} \int \D q\; e^{-\imag pq/\hbar}\psi(q,t=0)
  \\ & = \left( \frac{2\sigma^2}{\pi\hbar^2} \right)^{1/4} 
  \exp\left\{-\frac{\sigma^2}{\hbar^2}(p-p_0)^2 -\frac{\imag}{\hbar} q_0 p + 
    \frac{\imag}{\hbar} p_0 q_0 \right\} \,.
\end{split}
\end{equation}
Finally, we obtain the tomograms of the initial state using~(\ref{eq:tom}) as
\begin{equation}
  \tilde{w}(X,\mu,\nu,t=0) = \frac{1}{\sqrt{2\pi\sigma_T^2(\mu,\nu)}}\exp\left\{
    \frac{-(X-\mu q_0 -\nu p_0)}{2\sigma_T^2(\mu,\nu)}\right\}\,,
\end{equation}
where the width of the tomogram $\sigma_T$ depends on the particular
reference frame ($\mu,\nu$), and is given by 
\begin{equation}
  \sigma_T(\mu,\nu) = \sqrt{\nu^2\left(\frac{\hbar}{2\sigma}\right)^2 + \mu^2\sigma^2}\,.
\end{equation}

Throughout this work we express all quantities in terms of the 
fixed reference units for length ($u_\ell$), 
mass ($u_m$) and time ($u_t$).
From those we may also construct units for energy ($u_E=u_m u_\ell^2/u_t^2$)
and momentum ($u_p=u_m u_\ell / u_t$).
In these units $\hbar=u_m u_\ell^2/u_t$. 
To be specific, in (\ref{pot4}) we use $m=u_m$,
$\omega_0 u_t =0.1$, $a_3 u_\ell^2 = 0.01$, 
$\omega_4 u_t = 0.4$, $a_4 u_\ell^2 = 0.02$ and $V_0=u_E$.
As initial conditions, we choose $q_0=-5u_\ell$, $p_0=u_p$ and 
$\sigma=u_\ell/\sqrt{2}$.

\subsection{Discussion of the time evolution}
\paragraph*{Probability densities}
\begin{figure}
  \centering
  \includegraphics[width=\linewidth,clip]{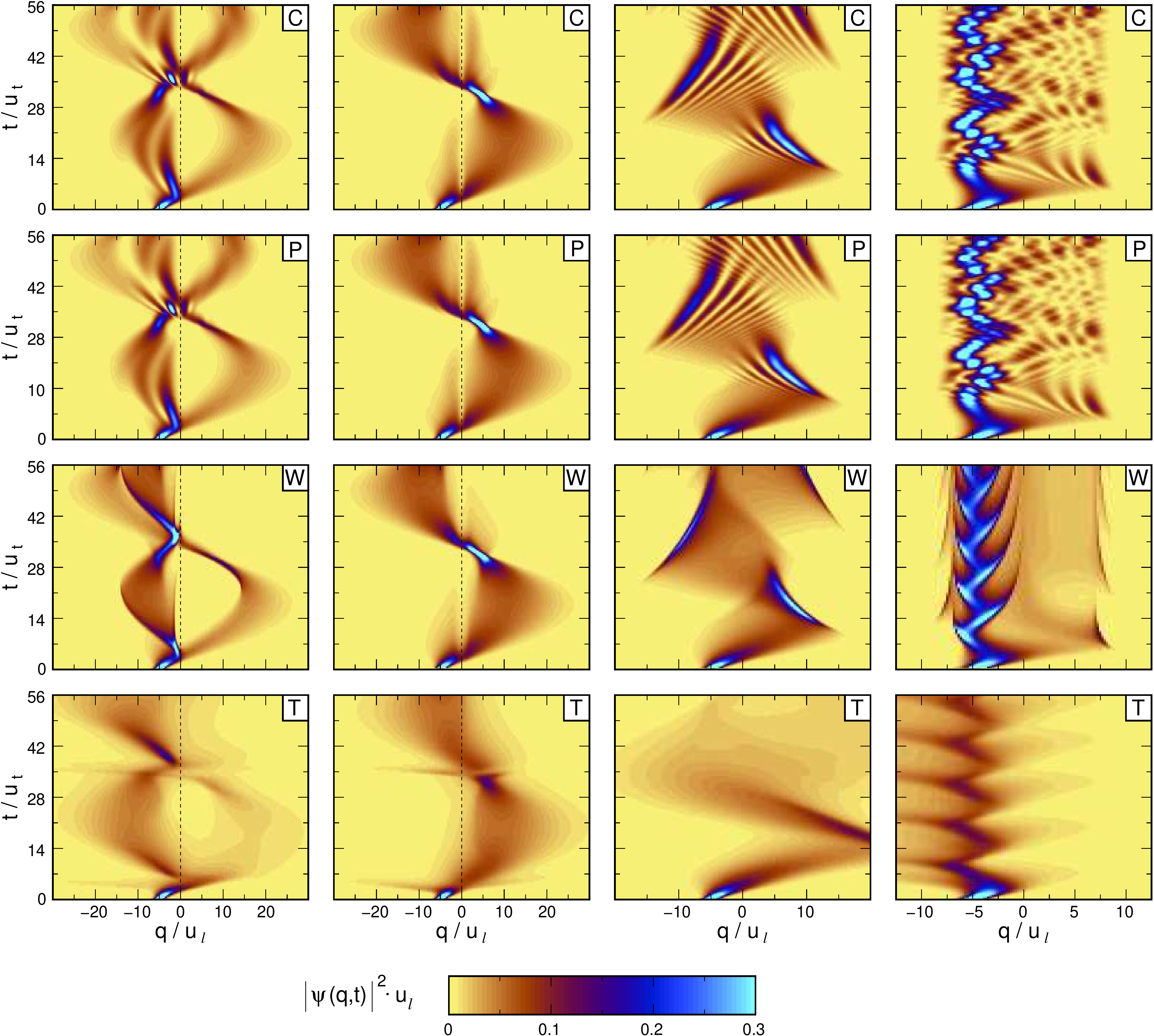}
  \caption{From left to right: Time evolution of the probability density
    in coordinate space, $|\psi(q,t)|^2$, for a particle in the 
    benchmark potentials $V_1(q)$ to $V_4(q)$ as given 
    by~(\ref{pot4}).
    For each case, we show (from top to bottom) the exact solution, 
    calculated by Chebyshev expansion ($C$), the results from 
    the linearized semiclassical propagator method ($P$), the results
    from the first order Wigner-Moyal approach ($W$) and the tomographic
    results ($T$). For implementation details see 
    Sect.~\ref{sect:num_aspects}.
    }
  \label{fig:overview}
\end{figure}
Figure~\ref{fig:overview} shows the time evolution of the probability 
density in coordinate space, $|\psi(q,t)|^2$, for times up to $t=56 u_t$.
Each column corresponds to one of the above potentials~(\ref{pot4})
while the rows refer to the methods used.
The topmost row, calculated by Chebyshev expansion ($C$), gives the exact
solution.
For the barrier case [$V_1(q)$, first column] the particle hits 
the barrier (center position marked by a dashed line), at about
$t\approx5 u_t$. 
Here, the main part of the wave packet is reflected, but a sizeable fraction 
also penetrates through the barrier.
The quantum nature of the particle gives rise to the interference pattern
on the left hand side of the barrier, where high and low probability densities
alternate.
This effect is still more pronounced around $t\approx 35 u_t$ when the 
transmitted and reflected parts interfere after their reunion.
For the case of a quantum well [$V_2(q)$, second column], we observe 
the overall picture expected for the dynamics in a simple harmonic trap.
Since the width of the initial Gaussian, however, does not match the resonance 
frequency of the harmonic trap, its width changes during the time evolution, 
refocusing once each $\pi/\omega_0$.
The rather small effect of the additional dip at $q=0$
consists in a reduced density in this region for all times and a retardation
of the transmission.
For the case of an anharmonic potential [$V_3(q)$, third column], 
besides the above mentioned broadening of the wave packet,
the nonlinearity of the force causes additional interference effects.
%
%
Elongations up to $q\approx 15 u_\ell$, exceeding the classical turning points 
$q_c\approx8.2u_\ell$, are possible due to the quantum nature of the 
initial state.
Despite its fixed total energy, the initial wave packet contains 
contributions with higher and lower energies.
The forth column shows the rich structure of the dynamics in the double
well potential.
While the principle part of the wave function remains in the left well, 
a considerable amount penetrates the barrier where strong interference 
effects arise.
The back scattering of these contributions causes strong interference 
patterns also in the left well for $t>10u_t$.

In the second row, the results from the linearized semiclassical propagator 
method $(P)$ fully coincide with the exact results.
The excellent agreement confirms, that within a semiclassical framework
it is in principle possible to capture quantum effects, although
by an tremendous increase of computational resources 
(cf. Sect.~\ref{sect:num_aspects}).

Restricting the iteration series for the Wigner function to first order, 
we obtain the results shown in the third row~$(W)$.
The initial splitting of the wave packet for the barrier case $V_1(q)$ 
at about $t\approx 5u_t$ is reproduced while the second one around 
$t\approx 35u_t$ is missed.
The explanation of this failure is simple.
Trajectories with high enough energy to cross the barrier once will 
be able to cross it any time, oscillating forth and back in the trap 
potential.
Others which already failed the first passage due to an insufficient energy
are reflected every time they hit the barrier.
Hence they stay on the left side of the barrier for all times and
in the second half period, $t> 35u_t$, there is almost no spectral
weight to the right of the barrier.
Nevertheless some, exponentially rare, trajectories can be found there.
Large negative initial momenta make some trajectories energetic enough to overcome
the barrier, but lead to a phase shift as compared to the trajectories
with positive initial momentum.
Furthermore, the first order Wigner results fail to resolve the fine 
structure of the interference effects for $V_3(q)$ and $V_4(q)$.
In this sense, the quantum information contained in this
approximation is limited even though in the initial Wigner function 
$W_0(q,p,t_0)$ all quantum effects (to arbitrary high orders of $\hbar$)
are included.
As we will see later, taking into account the second order term of 
the iteration series will alleviate these problems. 

The bottom line shows the results from the tomographic approach~$(T)$.
Here, the positions for the potential evaluation in 
(\ref{eq:TOM_prop_H}) are calculated from classical trajectories 
sampled from the initial distribution.
While also here, the dynamics of the system is described qualitatively, there
is nevertheless a large discrepancy to the exact data.
Four weak points should be stressed.
First, any sharp feature in $|\psi(q,t)|^2$ is washed out even more
than in the Wigner approach.
Second, for $V_1(q)$ and $V_2(q)$, the signatures at $t\approx5u_t$
erroneously extend to too large negative $q$-values.
The reason for this are diverging trajectories, caused by the negative
curvature of the barrier potential.
Third, the turning points in the anharmonic potentials ($V_3(q)$ and $V_4(q)$)
are highly overestimated, which is an effect of the coordinate sampling for the
potential evaluation.
Forth, the tunneling in the double well potential is not accounted 
for correctly.
This is due to the large range of $q$ for which the curvature 
of the potential is negative.
Hence, the amount of rejected trajectories is overestimated.
%

%
\paragraph*{Expectation values}

The semiclassical propagation of the Wigner function
and the tomogram is intented to give an adequate description of
complex many particle systems.
Thereby, average values take the center stage.

For the potentials $V_{\{1,2,4\}}(q)$ the initial wave packet splits in
two parts.
To account for this in terms of expectation values, we define
reduced average quantities for the positive and negative half-axis,
\begin{equation}
  \langle q\rangle_{-} = \frac{1}{N_{-}}\int\limits_{-\infty}^{0} 
  q | \psi(q,t)|^2 \D q \quad , \quad
  \langle q\rangle_{+} = \frac{1}{N_{+}}\int\limits_{0}^{\infty}
  q | \psi(q,t)|^2 \D q\,,
\end{equation}
where $N_{\pm}$ is the partial norm on the corresponding half-axis,
\begin{equation}
  N_{-} = \int\limits_{-\infty}^{0} | \psi(q,t)|^2 \D q \quad , \quad
  N_{+} = \int\limits_{0}^{\infty} | \psi(q,t)|^2 \D q\,.
\end{equation}

\begin{figure}
  \centering
  \includegraphics[width=0.9\linewidth,clip]{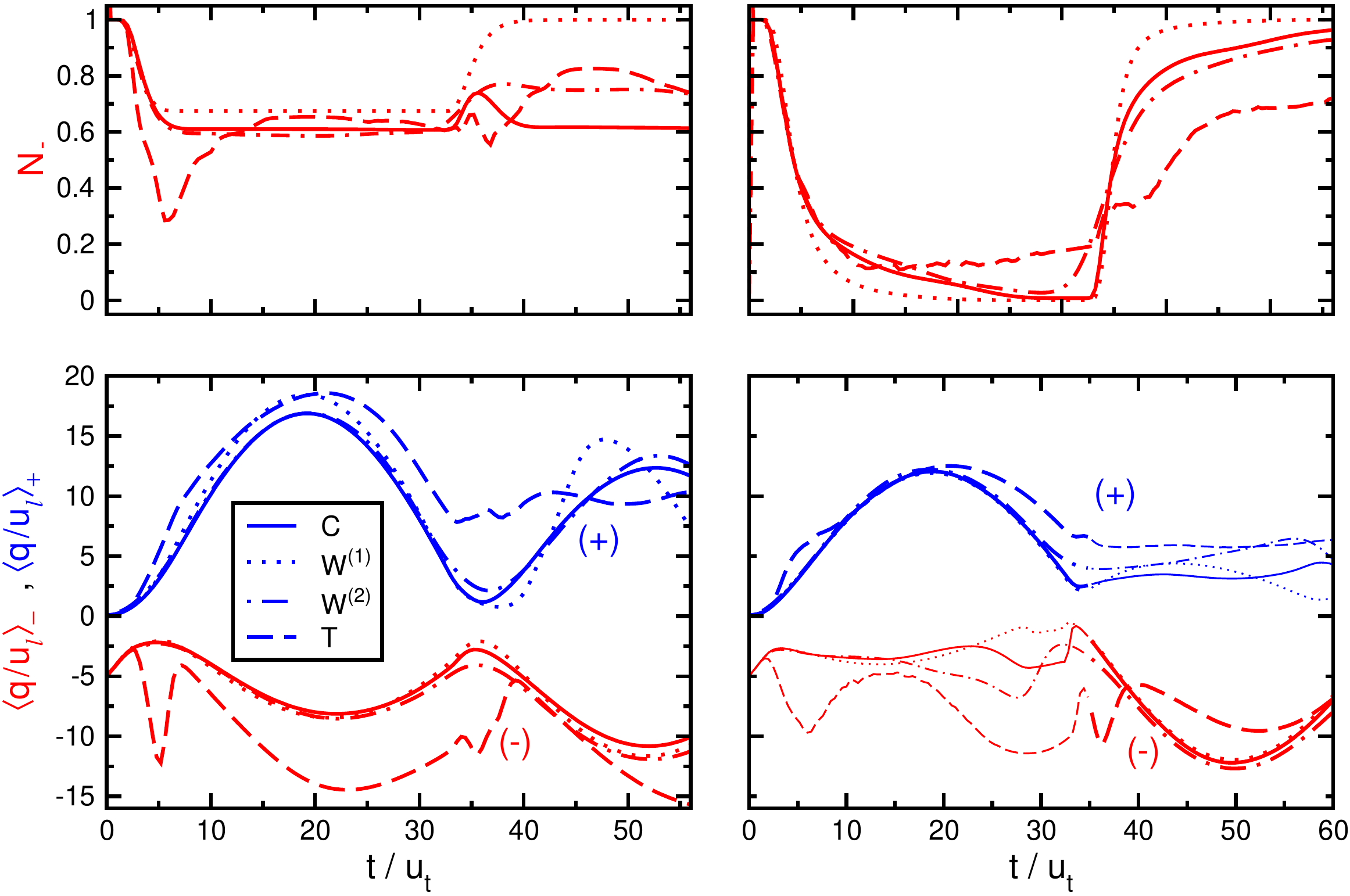}
  \caption{Expectation values for the benchmark potentials $V_1(q)$ (left column)
    and  $V_2(q)$ (right column). The upper panels show the part of the wave
    function norm on the negative half-axis $N_{-}$.
    In the lower panels, the mean positions $\langle q\rangle_{\pm}$ 
    on the corresponding half-axis are given.
    We show the exact results from the Chebyshev expansion ($C$ -- solid lines) 
    as well as the first and second order approximations from the Wigner-Moyal approach
    ($W^{(1)}$ -- dotted lines; $W^{(2)}$ -- dashed-dotted lines) 
    and tomographic results ($T$ -- dashed lines).
    As the wave functions obtained by the semiclassical propagator method agree 
    within numerical accuracy with the exact results from $C$, no separate curves
    are shown.}
  \label{fig:expectation_values}
\end{figure}

For the first two test cases, we show in Fig.~\ref{fig:expectation_values}
the time evolution of the initial state in terms of $\langle q\rangle_{\pm}$ and $N_{-}$.
The norm on the positive half axis is not shown separately as $N_{+}= 1- N_{-}$.
For $V_1(q)$, the constant value of $N_{-}$ after the initial tunneling event 
($t\approx 5u_t$) indicates an independent evolution of the two wave packets 
on their corresponding half-axis.
Only at the refocusing point ($t\approx 35u_t$) weight is temporarily 
redistributed between them due to interference effects.
From $\langle q\rangle_{\pm}$, the oscillation between the barrier and the 
confining trap of each wave packet can be identified.
The mean energy of the transmitted wave packet is larger
than for its reflected counterpart as the mean position reaches larger values.
Quantum dynamics, however, comprehends more than a simple 
energy discrimination of the constituents of the wave packet by the barrier. 

This becomes obvious when considering the first order Wigner result, $W^{(1)}$.
Containing the classical energy discrimination only, the
finite value of $N_+$ for $t> 35u_t$ is missed as already discussed in 
Fig.~\ref{fig:overview}.
The ratio of $N_+$ to $N_-$ within this approximation is solely determined by the 
initial energy of each simulated classical trajectory.
Those constituents of the initial state with $E>V_0$ will overcome 
the barrier while others will not.
For an estimate, we calculate the energy of the initial state,
neglecting for simplicity the influence of the barrier.
Then $N_-$ ($N_+$) is the integral over the initial Wigner function inside (outside)
an ellipse defined by $p^2/(2m)+\frac{1}{2}m\omega_0^2q^2=V_0$.
This result agrees well with the numerical data.
The strong underestimate of $N_{+}$ for $t> 35u_t$ within $W^{(1)}$ explains
the deviation of $\langle q\rangle_{+}$ in this range.

Taking into account the second order term of the iteration series, $W^{(2)}$, 
slightly improves the accuracies of both the norm and the reduced expectation values;
the relative weight of the split wave packet ($N_{-}$) after $t> 35u_t$ still 
deviates about 15\% from the exact value.
This deviation is related to the finite grid resolution, smearing out 
the exact positions during the recurrent coordinate sampling and grid deposition.
Increasing the grid resolution, this effect can be reduced systematically.

In the tomographic approach, the focus on expectation values does not resolve
the problems encountered for the probability density.
Around $t\approx 5u_t$, when mainly the barrier region is sampled, we 
observe two consequences of the negative potential curvature.
On the one hand, the fraction of trajectories which overcome the barrier is 
overestimated (too small value of $N_{-}$).
On the other hand, a considerable amount of the reflected trajectories 
diverges and thus the value of of $\langle q\rangle_{-}$ is too negative.
Apart from this the expectation values reproduce the results of $C$ 
qualitatively but not quantitatively.
Here the especially remarkable features are the larger elongation of 
$\langle q\rangle_{-}$ and the 
absence of a pronounced modulation of $\langle q\rangle_{+}$ for $t> 35u_t$.

For $V_2(q)$ (right column) the norm indicates the nearly perfect
transmission of the wave packet since for each time there is considerable 
weight on one half-axis only.
Therefore, the average positions show an almost perfect sinusoidal oscillation, if 
we consider $\langle q\rangle_{+}$ for $t<35 u_t$ and $\langle q\rangle_{-}$ afterwards
(thick parts of the solid lines).
On the respective other half-axis the average values should be taken with
care due to the low weight.
The rather good agreement between the expectation values from
$C$ and $W^{(1)}$ in this case is not surprising.
It is known from the literature that for harmonic potentials the exact Wigner 
function can be obtained by classical propagation of trajectories~\cite{Le95}.
As the dip around $q=0$ is only a moderate perturbation, $W^{(1)}$ describes
the dynamics still well.
For this case, also the method of Wigner trajectories~\cite{LS82,Le90b,Le92} is
applicable, in which the higher order terms of the iteration series are included
perturbatively into a pseudopotential $\tilde{V}(q)$.
The trajectories then follow the classical equations of motion with respect to
$\tilde{V}(q)$ instead of $V(q)$.
For the construction of such a pseudopotential an approximate Wigner function is
required. 
This limits a practical application of the method of Wigner trajectories to 
slightly perturbed harmonic potentials or nearly free motions~\cite{Le95}.
Furthermore, the existence of $\tilde{V}(q)$ is not guaranteed due to the 
perturbative character of this scheme.
Since for this potential the $W^{(1)}$ results for $\langle q\rangle_{\pm}$
almost perfectly match $C$ for the majority branch, the profit in considering
$W^{(2)}$ is limited.
Merely in the time evolution of $N_{-}$ an improvement from $W^{(1)}$ to 
$W^{(2)}$ is noticeable.
The tomographic results once more show a qualitative agreement with $C$.
In addition to the deviations for the minority branch, the diverging trajectories,
arising when the wave packet crosses the dip, perturb the results around
$t/u_t\approx 5,35$.

\begin{figure}
  \centering
  \includegraphics[width=\linewidth,clip]{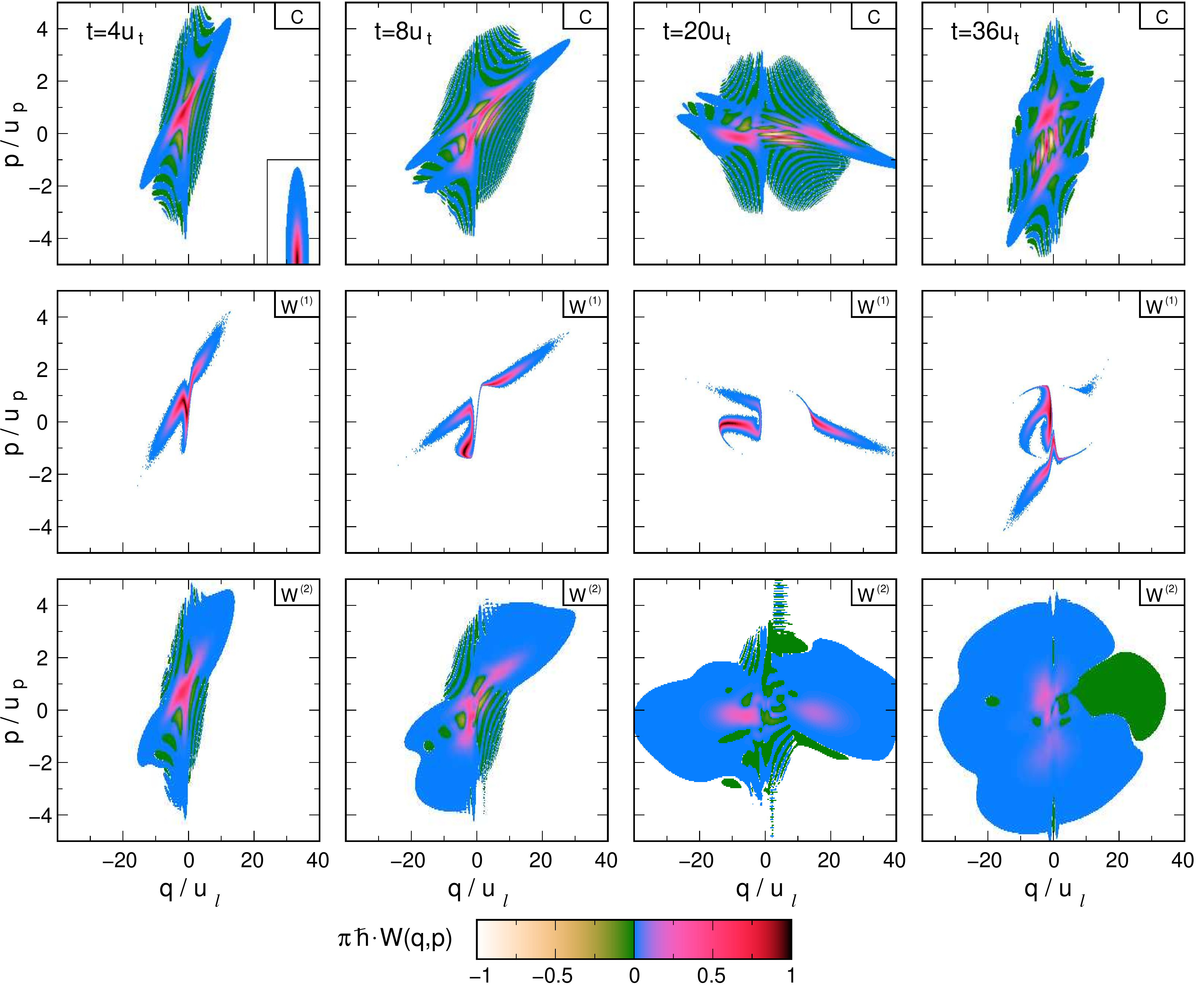}
  \caption{Time evolution of the Wigner function for potential $V_1(q)$.
    Upper panels:  Exact result using Chebyshev expansion ($C$). Middle panels:
    Propagation of the initial Wigner function using classical trajectories only
    [first order approximation, $W^{(1)}$].
    Lower panels: Inclusion of momentum jumps 
    [second order approximation, $W^{(2)}$]. 
    For each method snapshots at times 
    $t/u_t=4,8,20,36$ are shown. For comparison, the inset in the upper left
    panel shows the extension of a minimum uncertainty Gaussian.
  }
  \label{Wig_compare81}
\end{figure}

\paragraph*{Wigner function}

Focusing on the barrier case, $V_1(q)$, we compare in 
Fig.~\ref{Wig_compare81} the exact Wigner function, calculated from the
(Chebyshev propagated) wave function together with 
definition~(\ref{eq:Wigner}), to the first and second order approximation.
The extension of a minimum uncertainty Gaussian (inset in upper
left panel in Fig.~\ref{Wig_compare81}) is the lower limit where the 
concept of a joint probability holds due to the Heisenberg uncertainty relation.
Nevertheless, comparing $W(q,p)$ for different approximations can be taken as a good
quality check since any observable of the system is obtained as an
integral over the Wigner function.
Starting from the initial Wigner function (Gaussian, centered at $q=-5u_\ell$, 
$p=u_p$), the overall time evolution is dominated by a clockwise
rotation of the phase space points.
When the fastest contributions of the wave packet encounter the barrier
($t= 4u_t$), for the first time significant negative values of $W(q,p)$ occur.
The initial Gaussian shape breaks up into a triangle, reflecting the low-energy
trajectories which are held back by the barrier and the fast contributions 
overcoming the barrier.
The quantum nature is reflected in a weak interference pattern of small positive and
negative values around the main structure.
Evolving further in time ($t= 8u_t$), the regions with negative weights get 
more pronounced.
At $t= 20u_t$ there is a strong interference pattern of
large positive and negative values of $W(q,p)$ in between the two major
positive portions left and right of the barrier.
When the transmitted part returns to the left side of the barrier ($t= 36u_t$)
the structure remains divided with strong interference in between the two
positive bulks.

Considering the first order approximation, the most pronounced
difference compared to the full quantum result is the absence of regions
with negative values in $W^{(1)}$.
This is clear as the initial state $W(q,p,t=0)$ is strictly positive and during the
classical propagation of the trajectories their weight is unchanged.
Therefore, at any time $W^{(1)}$ is a superposition of positive 
contributions.
Despite the simplicity of this approximation, all regions with large positive 
weights are in essence captured correctly.
Regions of nearby positive and negative values in the exact solution
are marked within $W^{(1)}$ by vanishing, or strongly reduced values 
(cf. $q/u_{\ell}\in[0,10]$, $p/u_p\in[-1,1]$ at $t=20u_t$).
The integral over $p$ in this region vanishes for both $W^{(1)}$ and 
the exact solution, explaining why physically measurable
quantities like $|\psi(q,t)|^2$ agree well despite the differences in $W(q,p)$.

Including the second order term of the iteration series, the presence of negative
weights is restored in $W^{(2)}$.
Even though, arbitrarily large momentum jumps are allowed for the trajectories,
finite amplitudes of $W^{(2)}$ are restricted to the center region.
Outside this region, the fast oscillations of $\omega(s,q_{\tau})$ guarantee the 
complete cancellation present in the exact results. 
Taking the finite simulation grid into account, the extent of this cancellation
will delicately depend on the used resolution and accuracy of the (MC) integration.
A signature of a non-perfect cancellation can be seen for $t=20u_t$
at values of $q/u_{\ell}\in[5,10]$ and large $p$ as a series of
positive and negative stripes.
Accumulating such numerical fluctuations, the true region with finite 
amplitudes is overestimated for large times ($t=36u_t$).


\subsection{Details of implementation}
\label{sect:num_aspects}
%
%

\paragraph*{Chebyshev expansion}
In view of further applicability of the proposed methods, let us
focus on the computational requirements in what follows.
In this regard, the description of quantum dynamics by Chebyshev 
expansion of the time evolution operator is an extremely powerful method.
Its extraordinary performance is not restricted to the one-dimensional test cases 
considered here, but has been demonstrated successfully for more complex, higher
dimensional cases~\cite{SF08,FSSWFB09}.
%
%
Despite the exponential growth of the Hilbert space with the number of particles, 
also an application to many-particle systems, e.g., the polaron problem is
within reach~\cite{WF08}.
As compared to the standard Crank-Nicholson algorithm~\cite{PFTV86},
the Chebyshev expansion has two main advantages: speedup and larger 
accessible system sizes.
First, for the Chebyshev expansion only MVM are required,
while for the Crank-Nicholson scheme,
\begin{equation}
  \left(1+\frac{1}{2}\imag H \Delta t/\hbar \right) \; |\psi(t_0+\Delta t)\rangle = 
  \left(1-\frac{1}{2}\imag H \Delta t/\hbar \right) \; |\psi(t_0)\rangle\,,
\end{equation}
a linear equation system needs to be solved in each time step.
While for the one-dimensional case considered here, the tridiagonal structure
of the coefficient matrix speeds up the calculation, in general 
the solution of this system is the most time consuming step.
Speeding up the calculation by an initial inversion of the time independent 
coefficient matrix and performing successive MVM afterwards
is only feasible for moderate Hilbert space dimensions.

Second, the Crank-Nicholson algorithm is accurate to order $(\Delta t)^2$,
whereas the accuracy of the Chebyshev expansion is determined by the expansion
order $M$.
We may choose $M$ such that for $k>M$ the modulus of all expansion coefficients 
$|c_k(a\Delta t/\hbar)|\sim J_k(a\Delta t/\hbar)$ is smaller than a desired
accuracy cutoff.
This is facilitated by the fast asymptotic decay of the Bessel functions,
\begin{equation}
  J_k(a\Delta t/\hbar)
  \sim \frac{1}{\sqrt{2\pi k}} \left( \frac{\E a\Delta t}{2\hbar k} \right)^k 
  \quad {\mathrm {for}}\quad k\to \infty\;.
  \label{dec_Bessel}
\end{equation}
Then, for large $M$, the Chebyshev expansion can be considered as 
quasi-exact and thus permits a considerably larger time step.
For example, using a simulation grid of $N=1024$ sites with grid spacing $\Delta\check{q}=0.08u_\ell$ the
necessary scaling parameters for the four test cases are $a = 160,161,226,2307$,
and for $k>M=108,108,140,1028$ all $|c_{k}|<10^{-16}$.
These cutoffs ensure that the wave function is exact for the used time step 
$\Delta t=0.4u_t$ for all times.
Here, `exact' means that within numerical accuracy the wave function agrees
with the time dependent wave function obtained by a full diagonalization of 
the Hamiltonian,
\begin{equation}
  |\psi(t)\rangle = \sum\limits_{n=1}^{N} e^{-\imag E_n t/\hbar} 
  |n\rangle\langle n| \psi(t=0)\rangle\,,
\end{equation}
where $|n\rangle$ are the (time independent) eigenstates of the system and $E_n$
the corresponding eigenenergies.

Besides the high accuracy of the method, the linear scaling of 
computation time with both time step and Hilbert space dimension are 
promising in view of further applications to more complex systems.
Almost all computation time is spent in MVMs, 
which can be efficiently parallelized, allowing for a good speedup on 
parallel computers.

%
%
\paragraph*{Linearized semiclassical propagator method}
The considered implementation of the linearized semiclassical propagator 
method is not intended for a fast determination of an approximate
solution.
In this respect, more efficient flavors
of semiclassical propagator methods can be found in the literature, 
where higher order potential terms are taken into account and the propagated 
trajectories are chosen by some kind of importance sampling.
Instead, we focus on the question how close a semiclassical approximation
can be to the exact quantum solution if we let all concerns about 
computational requirements aside.
To achieve the desired accuracy, two aspects are of prominent importance.
First, we cannot choose the trajectories that have to be propagated  
solely according to the current wave function amplitude at the grid points.
For the overall interference effects, also trajectories starting at
grid points with low amplitudes are of importance.
Already discarding trajectories with initial weight $|\psi(q_0,t_0)|<10^{-6}$
influences the interference pattern and
leads to a noticeable deviation from the exact results in phase and
also magnitude.
Second, the used time step has to fulfill the Courant criterion~\cite{HE88}.
During a single time step any trajectory has to stay within its initial 
grid cell to ensure the validity of the local potential approximation.
Fortunately, the maximum distance a trajectory may cover in one time step
can be calculated exactly to optimize the time step.
From (\ref{eq:traj_q}) we read the displacement in one time step 
$\Delta t$ as $ \Delta q = -s(\Delta t)^2/(2m) + p_0 \Delta t / m$.
Substituting the largest value for $|p_0| = \pi\hbar/(\Delta \check{q})$ and 
requiring $|\Delta q| <\Delta \check{q}/2$, the maximum time step
is given by
\begin{equation}
  \Delta t_{\mathrm{max}} = \frac{\pi\hbar}{|s|\Delta \check{q}}\left(
    \sqrt{1+\frac{|s|(\Delta\check{q})^3 m}{\pi^2\hbar^2}}-1\right)\,.
\end{equation}
As for the reconstruction of the wave function contributions from the whole
grid are necessary, the grid points with the largest slope $s$ will be
most restrictive for the time step.
For $V_1(q)$ to $V_4(q)$ the used time steps are $\Delta t/u_t = 5\times10^{-3},
5\times10^{-3},5\times10^{-4}, 10^{-4}$ on grids with 
$N=1024,1024,512,256$ sites and grid spacing $\Delta\check{q}=0.125u_\ell$.
Using those parameters, the results reproduce within numerical accuracy
the results from the full quantum calculation, including the correct 
phase of the wave function.

%
%
\paragraph*{Wigner-Moyal-approach}
The numerical demands of directly propagating the Wigner function 
depend drastically on the order of the iteration
series taken into account.
For $W^{(1)}$, the classical propagation and assembly
of a large number of paths ($N_p\approx10^7$) is possible at very low
computational costs.
%
%
The continuity of the phase space trajectories allows for once sampling
the initial conditions and then following those paths up to arbitrary times.
In contrast, for higher order approximations a single initial sampling is not sufficient
anymore.
Due to momentum jumps also regions far away from the classical end points
acquire a finite weight.
Then, a resampling of the initial conditions at each time grid point is necessary.
Performing such a resampling also for $W^{(1)}$ slightly 
influences the data shown in Fig.~\ref{Wig_compare81}.
While for short times the agreement is almost perfect, with 
increasing time the sharp features present in the continuous
results are washed out by the resampling.
Caused by the grid discretization this effect systematically decreases with 
the used number of grid points.
The shape function used in the deposition to the grid influences
the necessary number of trajectories to achieve a desired accuracy.
Using the `cloud-in-cell' (CIC) scheme~\cite{HE88}, weight is attributed to 
the two nearest grid points in each direction, constituting a reasonable compromise
between deposition costs and necessary broadening.
For the first order results shown in 
Figs.~\ref{fig:overview}~-~\ref{Wig_compare81} we used
$N_p=10^7$ initially sampled trajectories.
Those are deposited  by the CIC scheme onto a $400\times400$ grid with 
$\Delta\check{q}=0.225u_\ell$ and $\Delta\check{p}=0.045u_p$.

\begin{figure}
  \centering
  \includegraphics[width=\linewidth,clip]{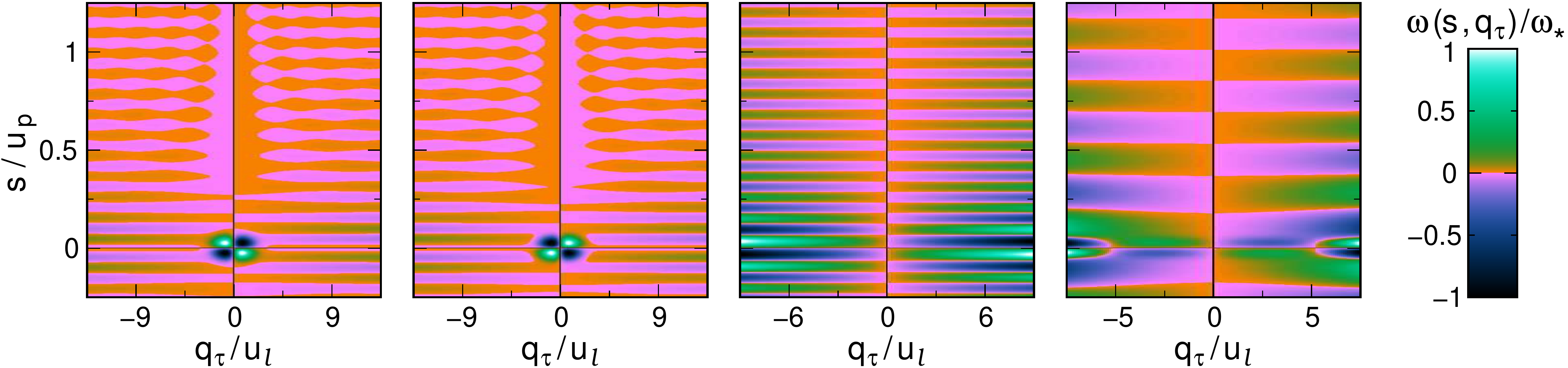}
  \caption{Weighting function $\omega(s,q_{\tau})$ in~(\ref{s6}) for the
    potentials $V_1(q)-V_4(q)$ (left to right). The shown functions
    $\omega(s,q_{\tau})$ are normalized to $\omega_\star u_m u_\ell = 300\;(450)$ for 
    the potentials $V_{\{1,2\}}(q)$ ( $V_{\{3,4\}}(q)$).
    The derivative of the $\delta$-distribution is implemented according to
    (\ref{eq:diff_delta}) with $\sigma_\delta=\Delta \check{p}=0.045u_p$.
    Note that the oscillation period in $s$ direction is influenced by the 
    range of considered $q_{\tau}$ and $q'$ in the integration~(\ref{s6}). 
    Here we used $|q'/u_\ell|,|q_{\tau}/u_\ell| <30,30,30,10$.
  }
  \label{fig:Omega_sq}
\end{figure}

For $W^{(2)}$, the computational 
requirements increase drastically.
While the recipe how to implement this order approximation is straight forward, 
improving the accuracy as compared to the first order results is
numerically challenging.
Especially the stability of the long time evolution depends drastically on 
several factors.
First, in each time step the MC sampling of the initial conditions depends 
on the previous result. 
Therefore, numerical and statistical fluctuations may amplify in runs 
with too poor statistics. 
As for the considered one-dimensional systems a direct integration is 
feasible, we refrain from the MC integration to circumvent possible 
convergence problems.
Adapting Fig.~\ref{fig:scetch_wig} to the full integration scheme, 
the considered initial phase space points $(q_0,p_0)$ cover the whole 
$(\check{q},\check{p})$ grid.
As long as $t-t_0$ is not too large, the $\tau$-integral can be evaluated 
by the midpoint rule.
In absence of momentum jumps the classical trajectory evolves continuously 
in $[t_0,t]$.
As compared to the dependency on the magnitude of the momentum jump occurring
at $\tau$ the influence of the exact jumping time $\tau$ in 
$[t_0,t]$ is of minor importance.
Working on a fine $s$ grid, at $\tau$ all those momentum jumps are performed
for which the final position $(q_{\tau},p_{\tau})$ does not exceed the 
simulation grid.
From the updated phase space point, the corresponding trajectories are 
then evolved up to time $t$, where they are deposited onto the 
$(\check{q},\check{p})$ grid by means of the CIC scheme.
Second, for the calculation of $\omega(s,q_{\tau})$ the derivative of the
 $\delta$-distribution in (\ref{eq:omega}) needs to be implemented numerically.
Approximating the $\delta$-distribution by a Gaussian of width $\sigma_{\delta}$, the
corresponding derivative reads
\begin{equation}
  \frac{d \delta}{ds} = \lim\limits_{\sigma_{\delta}\to0}\left[-\frac{s}
    {\sqrt{2\pi\sigma_{\delta}^3}}\exp\left(-\frac{s^2}{2\sigma_{\delta}^2}\right)\right]\,.
\label{eq:diff_delta}
\end{equation}
Taking into account the finite resolution of the simulation grid, a finite
value of $\sigma_{\delta}$ is necessary despite the desired limit $\sigma_{\delta}\to0$.
In the calculation we use $\sigma_{\delta}=\Delta\check{p}$ and choose the resolution 
of the momentum jump grid as $\Delta\check{s}=0.1\Delta\check{p}$ in order 
to resolve the structure of $d\delta/ds$ in  the $s$ integration.
A further reduction of $\sigma_{\delta}$ (and an according refinement of the 
momentum jump grid) does not increase the quality of the results.
In addition, the maximum allowed time step has to be severely reduced to prevent numerical 
instabilities due to the increased $\omega(s,q_{\tau})$ values in the 
vicinity of $s=0$.
For the relevant range of $s$ and $q_{\tau}$ the function $\omega(s,q_{\tau})$
is shown in Fig.~\ref{fig:Omega_sq} for $V_1(q)$ to $V_4(q)$.
Using the same phase space grid as for the first order approximation, the time step
$\Delta t = t-t_0=0.04u_t$ fulfills the stability requirements for the chosen
value of $\sigma_\delta=\Delta\check{p}$.

%
%

\paragraph*{Tomographic approach}

The crucial point of the tomographic approach is a suitable sampling of the 
potential landscape entering (\ref{eq:TOM_prop_H}).
A straight forward implementation of (\ref{eq:TOM_total}) suggests the consideration
of the whole coordinate axis for each time step and each trajectory $(X,\mu,\nu)$.
Then depositing each varied Gaussian 
onto the coordinate axis to obtain $|\psi(q,t)|^2$, such an 
implementation would be closely related to the concept of Sect.~\ref{sect:QPIC}.
The main difference between both methods is the more complicated propagation 
of the individual trajectories in the tomographic approach.
Using a local harmonic instead of a linear expansion of the potential and
including a diffusive term allows this method to use a larger time step.
But the computational overhead caused by the complexity of the calculation 
of each time step clearly outweighs this profit.

Exploiting the major advantage of the tomogram -- its positivity -- calls for 
choosing the used coordinates by a MC procedure instead.
A direct sampling of the coordinates according to the current probability density 
$|\psi(q,t)|^2$, however, fails completely to reproduce the exact results. 
Instead of the splitting, only a diffusive broadening of the initial wave 
packet is observed for $V_1(q)$.
Apparently no trajectories overcome the barrier as in (\ref{eq:TOM_prop_H}) only 
the repulsive force from the potential but not the actual momentum distribution is
taken into account.
To include the interplay between potential and momentum, 
for the data presented in Figs.~\ref{fig:overview} and~\ref{fig:expectation_values}
we evaluate the potential at the positions of simultaneously propagated 
auxiliary trajectories.
Starting from a phase space point $(q,p)$, sampled from the initial (quantum) state, 
they are classically propagated in time.
Note that the auxiliary trajectories determine the potential entering in
(\ref{eq:TOM_total}), but the evolved tomogram exerts no back action on them.
Thus $(X,\mu,\nu)$, and correspondingly the center of the varied Gaussian for
the deposition of $|\psi(q,t)|^2$, may significantly deviate from the auxiliary
trajectory, especially if the potential is evaluated in a region of negative curvature.
The diverging signatures around $t/u_t =5,35$ for $V_1(q)$ are due to
auxiliary trajectories with energies of almost exactly $V_0$.
Those stay in the negative curvature region in the vicinity of the potential maximum
for a long time.
There the evolution of 
$(X,\mu,\nu)$ is governed by hyperbolic functions in (\ref{eq:TOM_prop_H}).

Apart from this, the pronounced smoothing of the results in Fig.~\ref{fig:overview} 
is striking.
The broadening during the time evolution results from the dependency of 
the width of the deposited Gaussians on $(X,\mu,\nu)$, 
spoiling a good resolution.
On the other hand, approximate results are accessible with much less 
($N_p=12000$) trajectories than for the other discussed semiclassical methods.
Controlling the steadily increasing width of the Gaussians requires some kind
of restarting procedure in which the tomogram is resampled by Gaussians
of unit width after a certain time.
Performing such a resampling at each time grid point, we again approach the concept
behind Sect.~\ref{sect:QPIC}.
Declining such a restarting procedure in this work, the given results 
demonstrate the limitation of the tomographic approach 
when sampling the auxiliary trajectories only once.


\section{Conclusion}

In this work, we have compared different semiclassical approaches to quantum
mechanics regarding their numerical implementation and efficiency.
Focusing on the time evolution of a wave packet in one-dimensional 
quantum structures, we studied tunneling, interference and nonlinearity 
effects.
Results were obtained for the probability density and various expectation 
values and contrasted against the exact quantum mechanical solution, 
calculated by means of a Chebyshev expansion technique.
The Chebyshev method is fast, numerically stable and therefore perfectly 
suited to resolve the full dynamics of a quantum system.
Accessible system sizes are much larger than the ones that can be reached by other 
direct solution schemes of the time-dependent Schr{\"o}dinger equation 
(e.g., using the Crank-Nicholson algorithm or full diagonalization).

A brute force implementation of the Feynman path integral can be performed
by adapting a linearized semiclassical propagator method, where 
the inclusion of `all possible paths' is traced back to 
the set of possible initial conditions on a discrete coordinate and momentum grid.
Having in mind that within this approach the computation time scales as
$N^2N_t$, where $N$ ($N_t$) is the number of space (time) grid points,
the applicability to more complex systems is obviously limited.
Instead, the implementation should be considered as a `proof of principle'
that quantum effects are accounted for correctly if one takes into account
the complete superposition of the complex weighted trajectories 
within a local linear approximation of the potential.
Implementations going beyond this linear potential approximation
require a full inclusion of the monodromy matrix.
If one, along this line, correctly takes into account any phase jumps at
the focal points, the time step may be increased without loss of accuracy.
On the other hand, the neglect of some trajectories by the MC sampling 
procedure of the initial conditions leads to a systematic loss of accuracy.
Adopting a probabilistic point of view, the Wigner representation of quantum
mechanics offers an alternative approach to quantum dynamics. 
In the Wigner-Moyal scheme the Wigner function is propagated in time
according to an equation of motion, being equivalent to the von Neumann equation.
We transform this equation of motion into an integral equation which can be 
solved in terms of an iteration series.
Then, to leading (first) order, classical trajectories are propagated in 
time.
Thereby their initial conditions are sampled from the initial Wigner function.
Here, the low computational costs outweigh the loss of some aspects of
quantum dynamics.
Trying to improve the quality of the approximation by including the next (second) 
order term of the iteration series, we are faced with a tremendous increase
of computation time.
This is due to the necessity of considering a large number of trajectories, a 
high resolution of the phase space grid and a correspondingly small time step, 
in order to avoid the amplification of numerical fluctuations.
Thus, in order to get the exact quantum mechanical results, we need an even 
higher numerical effort than for the linearized semiclassical propagator method.
Contrasting the Wigner function results of both orders, the gain in accuracy 
for the second order scheme is only moderate such that the additional 
computational overhead seems not justified.
Hence, the complexity of the implementation and the ill posed numerics impede
the application of the higher order Wigner-Moyal approach to the description
of more complex systems.

Finally, the tomographic representation of quantum mechanics aims at 
describing quantum dynamics in terms of a positive semidefinite function.
The quantum tomogram can be interpreted as a set of Radon transformations of the Wigner
function.
Relating its time evolution to a diffusive Markov process, the dynamics of the
system is governed by a set of stochastic 
integral equations derived from the Kolmogorov equations for the tomogram.
In the calculation of the drift and diffusion coefficients for the stochastic 
differential equations, we used the harmonic propagator deduced from
a local, second order approximation of the potential.
The sampling of the potential landscape is the crucial point of this method
and strongly influences the quality of the data.
Evaluating the potential at the coordinates of classically evolving trajectories,
we reproduce the quantum results qualitatively but not quantitatively.
As compared to the other considered semiclassical approaches, the quality of the
data is poor, especially for potentials with distinct negative curvatures.
We expect a noticeable improvement of the results only if a more efficient sampling
of the coordinates for the potential evaluation can be found.
With respect to the computational costs, the tomographic approach is 
slightly more expensive than the Wigner-Moyal approach in first order approximation
but much faster than the linearized semiclassical propagator method or 
the Wigner-Moyal approach in second order.

To summarize, although the above analyzed semiclassical methods in principle 
capture all quantum effects, they largely differ in quality and required 
computational costs.
If one is interested in a method to include minor quantum corrections on top of
a classical description, the first order Wigner approach is best suited as it 
provides a reasonable compromise between accuracy and computation efficiency.
Of course, it is possible to reproduce the complete quantum mechanical solution 
at the expense of a dramatic increase of the computing resources.
In this respect, the linearized semiclassical propagator method is slightly
more efficient than the second order Wigner approach.
While more quantum effects should be included in the tomographic than in 
in the first order Wigner description, the expectation values obtained by the 
former approach are not as good as expected.
These aspects have to be kept in mind when applying the above methods to
the time evolution of more complex systems.

\subsection* {Acknowledgments}
This work was supported by the DFG through the research program SFB
TR 24, the Competence Network for Technical/Scientific High-Performance Computing 
in Bavaria (KONWIHR) and the Helmholtz-Gemeinschaft through COMAS.
We thank M.~Bonitz, D.E. Dauger, V.K. Decyk, A.V.~Filinov, V.E.~Fortov, 
P. Levashov and J. Tonge for helpful discussions.
The numerical calculations have been performed on the TeraFlop compute cluster 
at the Institute of Physics, Greifswald University.

\bibliography{./sfmsf09} 
\bibliographystyle{elsart-num}

\end{document}